\documentclass[useAMS,usenatbib]{mnras}
\usepackage{graphicx}
\usepackage{amssymb}
\usepackage{amsmath}
\usepackage{subfigure,float,verbatim,placeins,array, tabularx,tabulary,tabu,booktabs,xtab,makecell}
\usepackage{ifpdf}
\usepackage{threeparttable}
\usepackage{gensymb}
\usepackage{color}
\usepackage{lastpage}

\def\aj{AJ}
\def\apj{ApJ}
\def\aap{A\&A}
\def\apjl{ApJL}
\def\apjs{ApJS}

\def\mnras{MNRAS}

\def\prd{Physical Review D}
\def\na{}
\def\jcap{JCAP}

\def\nat{Nature}

\title[Fornax dSph GCs]
{A dwarf-dwarf merger and dark matter core as a solution to the Globular Cluster problems in the Fornax dSph}
\author[Leung et al.]
{Gigi Y.\,C. Leung$^{1}$\thanks{Email: leung@mpia.de}, Ryan Leaman$^{1}$, Glenn van de Ven$^{2}$ and Giuseppina Battaglia$^{3,4}$\\
$^{1}$Max-Planck Institut f\"ur Astronomie, K\"onigstuhl 17, D-69117 Heidelberg, Germany\\
$^{2}$Department of Astrophysics, University Vienna, T\"{u}rkenschanzstrasse 17, 1180 Wien, Austria\\
$^{3}$Instituto de Astrofisica de Canarias, C/Via Lactea, s/n, E-38205 La Laguna, Tenerife, Spain\\
$^{4}$Departamento de Astrofisica, Universidad de La Laguna, E-38206 La Laguna, Tenerife, Spain
}
\date{Accepted 2019. Received 2019; in original form 2019}

\pagerange{\pageref{firstpage}--\pageref{lastpage}} \pubyear{2019}
\hypersetup{draft}
\begin{document}

\label{firstpage}

\maketitle

\begin{abstract}
The five globular clusters (GCs) of the Fornax dSph are puzzling for two reasons; the mass in GCs is high with respect to the galaxy's old stellar mass, and their survival and large distance ($>$ 1\,kpc) is at odds with naive expectations of dynamical friction. We present here a semi-analytic model, simultaneously addressing both problems in a comprehensive evolutionary framework for Fornax.  Key to the model is inclusion of: 1) hydrodynamical constraints on the GC formation locations, 2) self-consistent velocity distribution functions in the dynamical friction calculations and 3) expansion of GC orbits due to a past dwarf-dwarf merger in the orbit integrations.  The latter is crucial for reconciling the dynamical survival of the clusters, and their chemical properties with respect to the Fornax field stars.  We find that in order for four of the GCs to survive at their observed projected location, a dark matter core of size $r_\mathrm{c}>$ 1.5\,kpc and a dwarf merger with dynamical mass ratio of 1:5 $\leq \eta \leq $1:2 with Fornax is required. We support the merger scenario by showing that aspects of the field star metallicity distribution function and anomalous chemical properties of GC5, are representative of a merging galaxy which is $\sim$1/3 less massive than Fornax.  Together the chemical and dynamical models suggest a scenario where three in-situ GCs in proto-Fornax were ejected to the outskirts during the merger, a GC4 formed during the merger at about 10\,Gyrs ago, with GC5 being brought in by the merging galaxy to Fornax.
\end{abstract}


\section{Introduction}
Under the standard cold dark matter (CDM) paradigm, cosmological and $N$-body dark matter only simulations have found that the density profiles of dark matter haloes tend to have cuspy profiles in the inner regions \citep[e.g.][]{nfw96}. The scale free nature of dark matter suggests that these typically cuspy NFW halos are characteristic of all bound structures in a $\Lambda$CDM universe. 

However, the predicted steep inner slope of the DM density profile is at odds with several observational constraints in low mass galaxies. Rotation curve decomposition of nearby spiral and dwarf galaxies \citep[e.g.][]{oh11, adams14, bro15} has found that the slowly rising rotation curves are better fit by DM profiles with shallower logarithmic slopes to their inner density profiles. Additional observations of stellar kinematics of dSphs generally suggest cored dark matter profiles \citep[e.g.][]{walk11, amo13, bur15, zhu16}. This is known as the `cusp-core problem'. 

There are two possible solutions to this problem. It has been shown that baryonic feedback can remove the central density cusps in CDM haloes of dwarf galaxies to produce central cores \citep[e.g.][]{pen12,pon12}. In particular, \citet{read16} show that such cores are have sizes related to the underlying stellar distribution in dwarf galaxies. On the other hand, existing alternative theories such as warm dark matter (WDM), Bose-Einstein condensate dark matter ($\psi$DM) and self-interacting dark matter (SIDM) predict shallow inner density slope even in pure DM haloes. The inner slope and the size of the dark matter core vary between different species of DM particles, between different DM particle masses within each species \citep[e.g.][]{lov14, sch14}, and in the case of SIDM, also between different DM interaction cross-sections \citep{kap16}. Precise observational constraints on the shape and the core size of dark matter haloes are therefore of strong importance.

The low-mass dwarf spheroidal galaxies (dSph) around the MW and M31 provide excellent test beds for the nature dark matter, as they are highly dark matter dominated objects (only after ultra-faint dwarf galaxies). Aforementioned methodologies, such as the rotation curve decomposition and dynamical modelling with stellar kinematics, have limitations such as uncertainties in stellar mass-to-light ratio and mass-anisotropy degeneracies. As an alternative, the survival of globular clusters (GCs) in dwarf galaxies has also been used as constraints on the dark matter halo shape, as the tidal forces and dynamical friction forces are sensitive to the total density that the GC sees. 

For example, \citet{amo17} suggested that the survival of low-density star clusters in Eridanus II and Andromeda XXV favours cored dark matter density profiles as a cuspy dark matter halo would exert too large a tidal force and hence disrupt the clusters. Together with considerations of dynamical friction and stellar evolution, \citet{con18} also found that the size and projected position of the low-density cluster in Eridanus II suggest a cored dark matter halo.

Of the classical dSphs, Fornax contains five GCs, which, together with extensive ancillary data of the host (e.g. stellar velocity, age and metallicity measurements), makes it a unique test case for probing the nature of its dark matter halo. All but one of the five GCs have stellar masses of $>10^5\,M_\odot$, the massive GCs in Fornax are therefore not subjected to destruction by the tidal field of the host galaxy, unlike the GCs in Eridanus II. The large projected distances of 240 to 1600\,pc between the GCs and the center of Fornax, however, pose another challenge. With ages $>10$\,Gyr, they are naively expected to have already been brought to the center of the galaxy via dynamical friction from the field stars and the dark matter halo to form a nuclear star cluster \citep[e.g.][]{tre76, her98}. This is known as the `Fornax timing problem'. This discrepancy poses a challenge to our understanding of not only the $N$-body problem, but also the nature and structure of dark matter. 

$N$-body simulations have shown that the shape of the density profile of the underlying background particles has a profound impact on the orbital decay trajectory and therefore the time it takes for a massive infalling object to reach the galactic center \citep[e.g.][]{read06, in09, in11, cole11}. In particular, cored dark matter halo profiles are found to allow slower decay than cuspy halo profiles. In addition, the orbital decay is found to stall in cored halo profiles, before the massive infalling object reaches the galactic center. 

Semi-analytic prescriptions for dynamical friction \citep[e.g.][]{chan43} have shown some success at reproducing the orbital decay of a massive object under a background particle distribution. Several works have studied and verified the orbital decay of massive object under background particles of various density profiles. Notably, \citet{petts15} and \citet{petts16} have successfully reproduced the slower decay and the core-stalling effect of cored halo profile with the inclusion of tidal stalling and by adopting more a radially varying impact parameters. With detailed treatments of dynamical friction, the timing problem can therefore provide a constraint on the dark matter halo profile and hence allow a glimpse into the nature of dark matter.

Several solutions to the Fornax timing problem have been proposed in the literature. \cite{oh00} suggested that the survival of GCs in Fornax can be resolved by invoking massive black holes which scatter the GCs to large radii, or a strong external tidal field from the Milky Way. There is however a lack of evidence for the existence of such black holes in Fornax. More problematic is that the proper motion of Fornax suggests that the dSph had never been closer to the Milky Way than its present location \citep{lux10, hel18}, implying that Fornax had never encountered a sufficient tidal field from the Milky Way to expand its GCs' orbit to their observed locations. 

With $N$-body  simulations of the Fornax system, it has been shown that the GCs in Fornax would not reach the galactic center within a Hubble time with a cored profile \citep{goe06, read06, cole12}. \cite{cole12} have also reported a 'dynamical buoyancy' in their $N$-body simulations of the five GCs in Fornax orbiting in a dark matter halo with a core radius $r_\mathrm{c}$ of 1000\,pc. Such dynamical buoyancy would act as a force that pushes the GCs outwards, acting against the dynamical friction. While \cite{cole12} have performed the $N$-body simulations on four different halo profiles, only the profile with a large core shows noticeable dynamical buoyancy. With such a profile, two out of the five GCs can survive outside of the observed galactocentric distance. Interestingly, \cite{san06} have also ruled out MOND using the Fornax timing problem, as the GCs would fall into the galactic center too quickly ($\sim$1\,Gyr) under MOND. Conversely, \cite{hui17} show that dynamical friction would be largely reduced if dark matter is made up of the $\psi$DM superfluid. In addition to the cored density profile of $\psi$DM, the wave nature of $\psi$DM would suppress the over-densities formed behind the infalling GCs, leading to a weaker dynamical friction.

Without a constraint on the starting position of the GC's initial orbit, modelling the orbital decay due to dynamical friction provides an incomplete and unconstrained picture of the GCs history and origin of their present-day location. Previous studies therefore either focus on whether the dynamical friction timescale is larger than the age of the GCs \citep[e.g.][]{goe06, san06, hui17}, or reproducing the observed distance by forcing the GCs to be formed at $>$1000\,pc or even at the current tidal radius ($\sim$2000\,pc) \citep[e.g.][]{an09, cos09, arca16}. Given the measured age of the GCs (10-13\,Gyrs), it is unclear whether the gas density would have been high enough at those redshifts to support the formation of the GCs at such large galactocentric distance - especially given the more compact size expected for the high-redshift progenitor of Fornax. It is therefore crucial to incorporate gaseous and stellar disk evolution models when estimating the galactocentric distances at which the GCs are formed, when addressing the present day position of the GCs. 

For example, \citet{kru15} suggested that once formed in a central high (local) gas density environment, the GCs have to be ejected out of their formation environment to avoid disruption due to the strong chaotic tidal field of the gaseous interstellar medium. Such an ejection could be caused by dynamical interactions with gas clumps, stellar feedback or a merger. Past merger events might be expected in dSphs like Fornax as they are found also as a possible pathway for the transformation of gas-poor dSphs from gas-rich dwarf irregulars in cosmological simulations \citep[e.g.][]{wet15}. 

Specifically to Fornax, past merger events have been suggested in order to account for its complex metallicity distribution function, multiple stellar populations and differential internal dynamics between the populations \citep[e.g.][]{walk09,amo12}. In particular the younger stellar population has a rotation axis offset from the main rotation axis, and this kinematic misalignment could be the result of a merger. The long-standing star formation history (SFH) of Fornax may also be suggesting accretion of gas or stars from other systems \citep{delp13}. The large total mass of the GCs relative to the mass of metal poor field stars would also be alleviated if one or more of the GCs are accreted via a merger as pointed out by \cite{lar12}.

It is therefore crucial to incorporate possible influences on the positions of the GCs in Fornax due to the past merger event. Depending on the nature of the merger, the orbit of the GCs can undergo either an expansion or a contraction \citep[e.g.][]{naab09}. For a non-dissipative (dry) merger, the GCs' orbits would gain energy from the merger and expand according to the mass ratio between the host and the merging galaxies.

An additional aspect contributing to the orbital evolution which has been neglected in previous semi-analytic models of the GCs' orbital trajectories in Fornax is the aforementioned dynamical buoyancy as reported in \citet{cole12}. The dynamical buoyancy effect is particularly crucial if the formation location of the GC was at a galactocentric distance  less than the current day location. Clearly a holistic approach which takes into account, dynamical buoyancy/friction, along with physical constraints on the formation position and merger history of Fornax is necessary to provide a better understanding of the evolution of this unique galaxy. This would require an exploration of a wide range of halo profiles and merger mass ratios, which can be too computationally expensive to be done with $N$-body simulations. 

The goal of this work is to build the first semi-analytical model that includes the aforementioned ingredients: (1) a physically-motivated formation location of the GCs, (2) the effect of dynamical buoyancy and (3) a past merger, and then infer the underlying dark matter halo profile of Fornax by requiring the modelled current galactocentric distances of the five GCs to be outside of their observed projected distance ($d_\mathrm{p}$; see Table \ref{tab_gc}). In the following sections, we first present the ingredients of our semi-analytical model in Section \ref{sect_model}. This includes the estimation of the formation location of the GCs, the density profiles of background dark matter and stellar particles, the dynamical friction treatment with dynamical buoyancy implementation, and orbital expansions caused by mergers. We then present the result in Section \ref{sect_results}, which is followed by a discussion on how the dark matter halo parameters we obtained compare with respect to $\Lambda$CDM cosmological simulations and whether the required merger mass ratio in our model is consistent with the observed metallicity distribution function in Section \ref{sect_diss}. We summarise our key findings and conclude in Section \ref{sect_con}.

\section{Semi-analytic Model}\label{sect_model}
\begin{figure*}
\begin{center}
\includegraphics[width=0.9\textwidth,clip = True, trim=0 20 80 0]{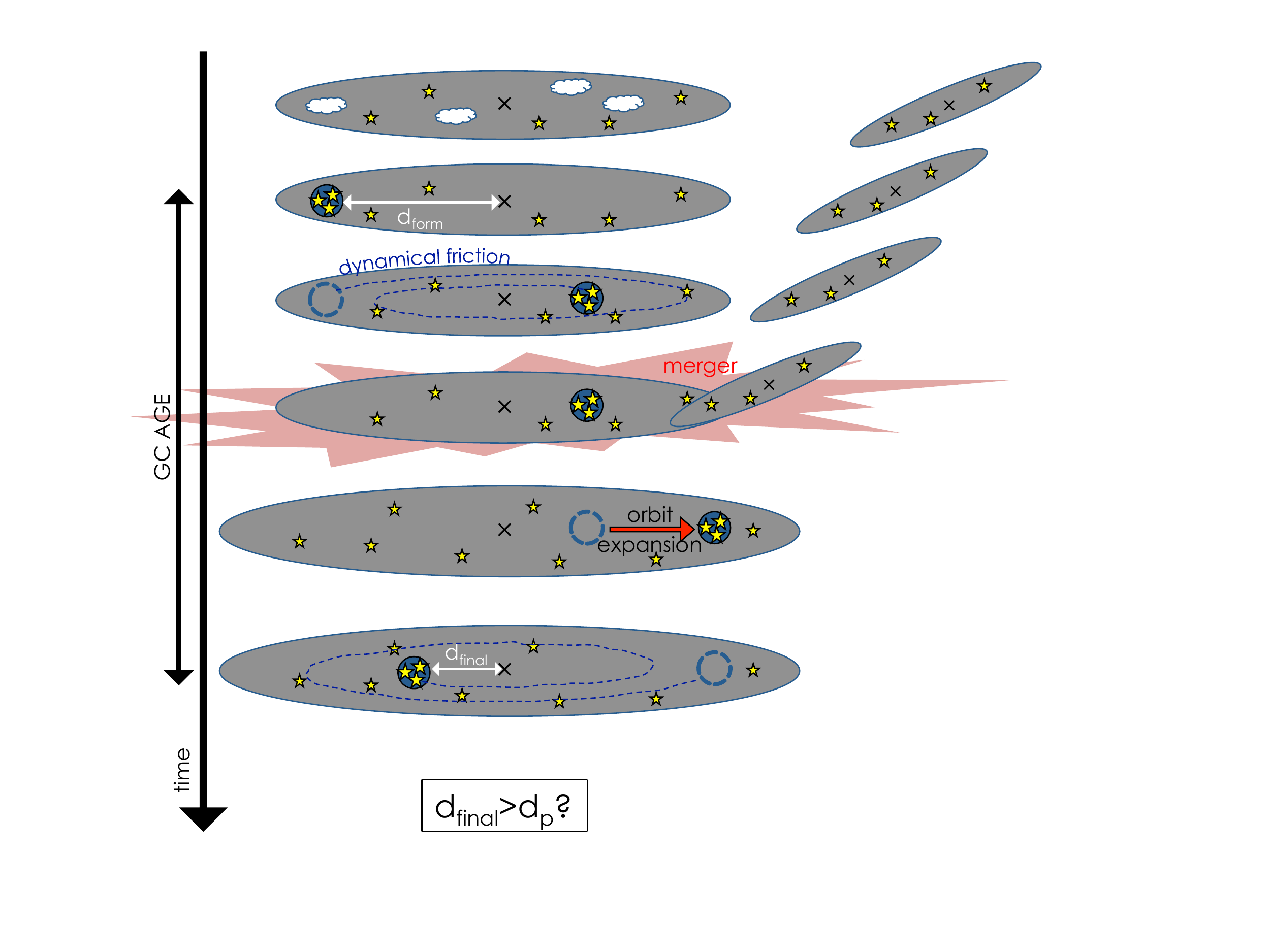}
\caption{Schematic diagram of our semi-analytical model. We first constrain the formation location of the GCs ($d_\mathrm{form}$) using a pressure equilibrium argument. The GCs' orbits then undergo dynamical friction, calculated for each of the specified dark matter halo profiles, throughout their lifetime (i.e. ages). In between, a merger with a companion of a specified mass ratio is incorporated which causes orbit expansions to the GCs. The modelled present-day positions of the GCs ($d_\mathrm{final}$) are then compared to the observed projected distance ($d_\mathrm{p}$) to constrain the dark matter halo profile.}
\label{fig_model}
\end{center}
\end{figure*}

In the following section we will describe the ingredients that go into building our semi-analytic model of the co-evolution of Fornax and its GCs.  The model is unique in that it provides physically motivated expressions for the GC formation distance, an updated dynamical friction/buoyancy prescription and the effect of a dwarf-dwarf merger on the orbits of the GCs. A schematic representation of all the ingredients of our model can be found in Figure \ref{fig_model}. We should emphasise here that while our model calculates three-dimensional final galactocentric distances ($d_\mathrm{final}$) of the GCs, the lack of precise measurements of the line-of-sight distances of the GCs means that we are not able to project the three-dimensional $d_\mathrm{final}$ to fit directly to the observed $d_\mathrm{p}$. Instead, we conservatively require $d_\mathrm{final}>d_\mathrm{p}$ in providing a lower-bound to the DM core size and the merger mass-ratio.

\subsection{Constructing the host galaxy Fornax}\label{subsect_bg}
We represent Fornax with two components: a dark matter halo and a spherical stellar distribution. The dark matter halo is parametrised by a `cored NFW' profile (cNFW), which was found to be a good description of simulated dark matter haloes on dwarf galaxies which were altered by baryonic feedback mechanisms \citep{read16}:
\begin{equation}
\begin{aligned}
&\rho_\mathrm{NFW}(r)=\rho_0\Big(\frac{r}{r_s}\Big)^{-1}\Big(1+\frac{r}{r_s}\Big)^{-2} \\
&\rho_\mathrm{cNFW}(r)=f^n\rho_\mathrm{NFW}+\frac{nf^{n-1}(1-f^2)}{4\pi r^2r_c}M_\mathrm{NFW} \\
&M_\mathrm{cNFW}=M_\mathrm{NFW}f^n,
\end{aligned}
\end{equation}
where $\rho_0$ is the characteristic density, $r_s$ is the scale radius, $r_c$ is the core radius, $\rho$ and $M$ represent the density and enclosed mass profile of the respective halo, $f^n$ renders the profile at $r<r_{c}$ to be shallower than an NFW profile and can be written as:
\begin{equation}
f^n=\Big[\tanh\Big(\frac{r}{r_c}\Big)\Big]^n,
\label{eq_read}
\end{equation}
and $n$ is a parametrisation of how 'cored' a profile is with $n=0$ representing an NFW profile and $n=1$ representing a completely cored profile. In this work we test the limiting case of $n=1$ for all dark matter profiles.

\begin{figure}
\begin{center}
\includegraphics[width=0.47\textwidth,trim=28 10 397 25,clip=True]{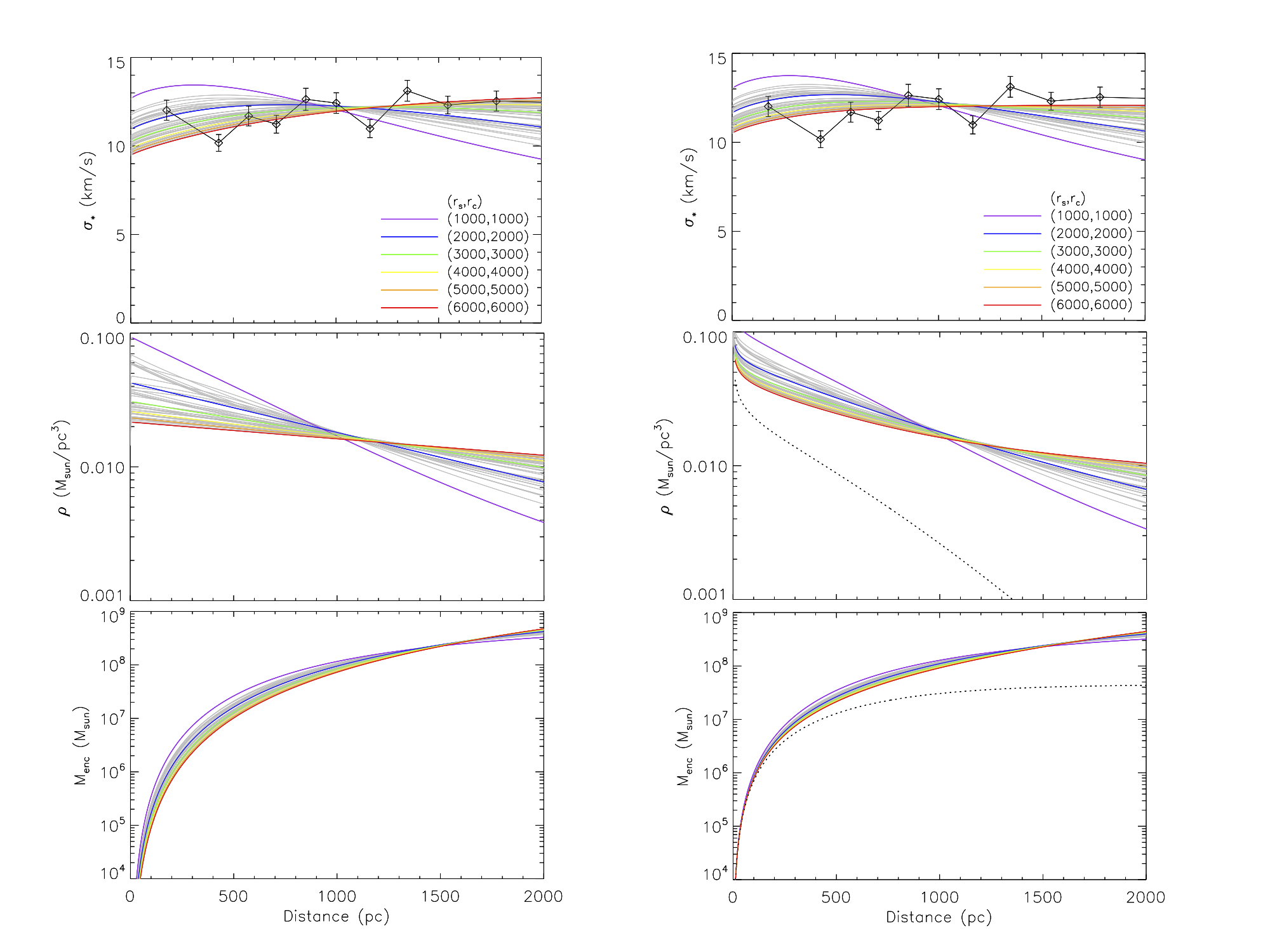}
\caption{\textit{Top:} the observed stellar velocity dispersion radial profile of Fornax \citep{bat06} is plotted in black diamonds with error bars. Overlaid in grey are all the dark matter profiles we tested in our ($r_\mathrm{s} ,r_\mathrm{c}$) grid. We show in colour six examples of the $\sigma_\star$ profiles from our normalised dark matter profiles. \textit{Middle and bottom:} the corresponding density and enclosed mass profiles. 
}
\label{fig_sigma}
\end{center}
\end{figure}

The surface brightness profile of the stellar component, $\Sigma_\star(R)$, is described using a Sersic profile as fitted by \citet{bat06}:
\begin{equation}
\Sigma_\star(R) = \Sigma_{0, \star}\exp\Big[\Big(\frac{R}{R_\mathrm{s}}\Big)^{1/m}\Big],
\label{eq_sersic}
\end{equation}
where $R$ is the 2D-projected radius, $R_\mathrm{s}$\,=\,694.5\,pc, $m$\,=\,0.71, and $\Sigma_{0, \star}$ is obtained through a normalisation to the total stellar mass in Fornax of 4.3$\times$10$^{7}$\,M$_\odot$ \citep{deB12}. The surface brightness profile is then deprojected to a density profile $\rho_\star(r)$ using Eq.\,17-19 of \citet{lim99}. The density profile of the stellar component takes the following form throughout this paper:
\begin{equation}
\rho_\star(r) = \rho_{0, \star}\Big(\frac{r}{R_\mathrm{s}}\Big)^{-p}\exp\Big[-\Big(\frac{r}{R_\mathrm{s}}\Big)^{1/m}\Big],
\end{equation}
with $\rho_{0, \star}$\,=\,0.015\,M$_\odot$\,pc$^{-3}$ and $p$\,=\,0.25\footnote{Here $\rho_{0, \star}$ is obtained again through the normalisation of the total stellar mass and $p$ is given as a function of $m$ (as defined in Eq. \ref{eq_sersic}) in \cite{lim99}.}. The stellar density and enclosed mass profiles are plotted in the middle and bottom panel of Figure \ref{fig_sigma} respectively.

\begin{figure}
\begin{center}
\includegraphics[width=0.53\textwidth,trim=75 0 145 0,clip=True]{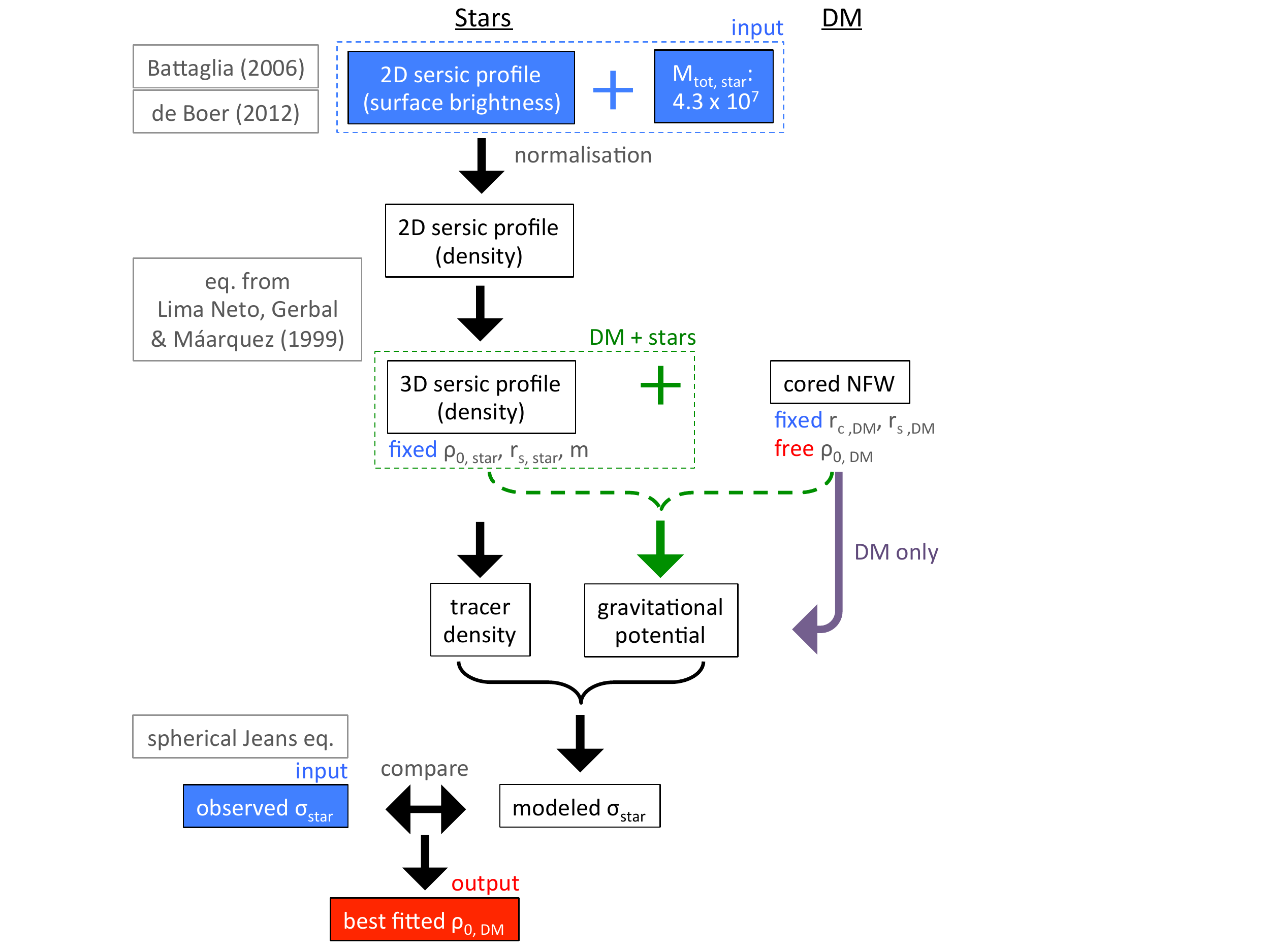}
\caption{A flow chart showing the steps for normalising dark matter halo profiles with various ($r_\mathrm{s}, r_\mathrm{c}$). The purple and green paths show how the gravitational potential is obtained in the dark matter only and the dark matter + stars case respectively. }
\label{fig_dmnorm}
\end{center}
\end{figure}

While the density distribution of the stellar component is fixed in our semi-analytic model, the $r_\mathrm{s}$ and $r_\mathrm{c}$ of the dark matter halo remain as free parameters. For each ($r_\mathrm{s}$, $r_\mathrm{c}$), the corresponding $\rho_0$ is obtained through a normalisation to the observed stellar velocity dispersion $\sigma_\star(R)$ from \cite{bat06}. The $\sigma_\star(R)$ for each halo profile is estimated with the Jeans equation under the spherical and isotropic assumption:
\begin{equation}
\label{eq_jeans}
\sigma^{2}(r)=\frac{1}{\rho(r)}\int_r^{\infty} \rho(r)\frac{d\Phi}{dr'}dr',
\end{equation}
where $\sigma(r)$ and $\rho(r)$ in this case is the intrinsic velocity dispersion and density profile of the tracer particle, i.e. $\sigma_\star(r)$ and $\rho_\star(r)$ and $\Phi$ is the corresponding gravitational potential computed from the density distribution of the background particles (dark matter and/or stars). The binned stellar velocity dispersion for each of the total potentials is shown in Figure \ref{fig_sigma}. As examples we over-plotted the $\sigma_\star(R)$ (obtained through the 2D-projection of $\sigma_\star(r)$)\footnote{Once again, $r$ and $R$ here provide the distinction between the 3D and 2D projected radii respectively.} of six different profiles in Figure \ref{fig_sigma}. We summarise our steps in normalising the dark matter halo profiles in Figure \ref{fig_dmnorm}. This is not an attempt to get a `best-fit' dark matter profile from the observed $\sigma_\star(R)$ profile, but rather, to illustrate the degeneracies between various profiles when using just the observed $\sigma_\star(R)$ as a constraint and to show that the normalisation of our dark matter halo profiles are reasonable.

\subsection{Constraining the formation location of the Globular Clusters}\label{subsect_rstart}
While the detailed formation physics of dense star clusters is currently debated, there are simple analytic estimates for the necessary environment of the gaseous regions which they are expected to form from. In particular, \citet{elm97} suggested that the star clusters kinematic density may form in pressure equilibrium with the mid-plane pressure of the surrounding molecular gas phase. To constrain the starting location of the GCs, we consider a pressure equilibrium scenario at their formation. In such scenario, the external pressure of the galactic disk ($P_\mathrm{ext}$) should be equal to the internal pressure of the GC ($P_\mathrm{in}$) itself. $P_\mathrm{in}$ can be written as: 
\begin{equation}
P_\mathrm{in} = 4\pi G\Sigma_\mathrm{GC}^{2} = 4\pi G(\frac{M_\mathrm{GC}}{\pi R_\mathrm{GC}^{2}})^{2},
\end{equation}
where $G$ is the gravitational constant, $\Sigma_\mathrm{GC}$, $M_\mathrm{GC}$ and $R_\mathrm{GC}$ are the surface density, mass and half-mass radius of the GC respectively. $M_\mathrm{GC}$ and $R_\mathrm{GC}$ are listed in Table \ref{tab_gc}. $P_\mathrm{ext}$ is related to the gas surface density ($\Sigma_\mathrm{gas}$), stellar surface density ($\Sigma_\star$) and the ratio between the velocity dispersion of gas and star ($f_\sigma=\sigma_\mathrm{gas}/\sigma_\star$) by:
\begin{equation}\label{eq_pext}
P_\mathrm{ext} = 4\pi G\frac{\pi}{2}\Sigma_\mathrm{gas}(\Sigma_\mathrm{gas}+f_\sigma\Sigma_\star).
\end{equation}

\begin{table}
\begin{tabular}{@{}lccccc}
\hline

GCs & $M_\mathrm{GC}$ &Age & [Fe/H] &$d_\mathrm{p}$ & $R_\mathrm{GC}$\\
& ($10^{5}\,M_{\odot}$) &(Gyr) & &(pc) &  (pc)\\ 
\hline
\hline
GC1 & 0.37 & $12.1\pm0.8$ & $-2.5\pm0.3$ & 1600 & 10.03\\
GC2 & 1.82 & $12.2\pm1.0$ & $-2.5\pm0.3$ & 1050 & 5.81\\
GC3 & 3.63 & $12.3\pm1.4$ & $-2.5\pm0.2$ & 430 & 1.60\\
GC4 & 1.32 & $10.2\pm1.2$ & $-1.2\pm0.2$ & 240 & 1.75\\
GC5 & 1.78 & $11.5\pm1.5 $ & $-1.7\pm0.3$ & 1430 & 1.38\\
\hline
\end{tabular}
\caption{Properties of the five globular clusters of Fornax dSph. The masses ($M_\mathrm{GC}$) are taken from \citet{mac03m}\protect\footnotemark. The ages are taken from \citet{deB16}. Metallicities are taken from \citet{deB16}. The projected distances ($d_\mathrm{p}$) of GC1, 2, 3 and 5 are determined with the central position of the five GCs determined by \citet{mac03m}. 
The radii of the GCs ($R_\mathrm{GC}$) listed here are the fitted core radii of a King model from \citet{mac03m}. }
\label{tab_gc}
\end{table}
\footnotetext{While newer measurements of the GC masses are also available from \cite{deB16}, the two measurements are consistent within the uncertainties. We chose the older value from \cite{mac03m} for easier comparison with other works on the dynamical friction of GCs in Fornax. In particular, we would like to reproduce the effect of dynamical buoyancy, and compare the resultant stalling radii with that of \cite{cole12} using the GC mass measurements from \cite{mac03m}.}

To obtain the gas and stellar surface density at the formation epoch of the GCs, we utilise the star formation history (SFH) of Fornax dSph. We obtained the SFH from \citet{deB12}. We then assume that the star formation rate (SFR) has an exponential profile with radius at any time epoch and create SFR profiles $\Sigma_\mathrm{SFR}$($R$) for $t=t_\mathrm{GC}$, where $t_\mathrm{GC}$ is the age of the globular cluster. 
From $\Sigma_\mathrm{SFR}$($R$) we can obtain the gas surface density of the disk $\Sigma_\mathrm{gas, disk}(R)$ by adopting a depletion timescale $\tau_\mathrm{dep}$ such that $\Sigma_\mathrm{gas, disk}(R)=\tau_\mathrm{dep}\Sigma_\mathrm{SFR}(R)$. We adopt the cosmological model from \cite{dut09} to allow the scale radius of the exponential profile of the gaseous disk to grow with time. To account for the fact that GCs often form in overdense regions of giant molecular clouds, the final $\Sigma_\mathrm{gas}$ we adopt for Equation \ref{eq_pext} is \citep{kru15,kru05}:
\begin{equation}
\Sigma_\mathrm{gas} = 3.92\Sigma_\mathrm{gas,disk}(5-4(1.+0.025(\Sigma_\mathrm{gas, disk}/100)^{-2})^{-1})^{1/2}
\end{equation}
The stellar surface density profile $\Sigma_\star(R)$ is then obtained by integrating the SFH from $t=13.6\,\mathrm{Gyr}$ to $t=t_\mathrm{GC}$.

\begin{figure}
\begin{center}
\includegraphics[width=0.5\textwidth,clip = True, trim=10 35 20 80]{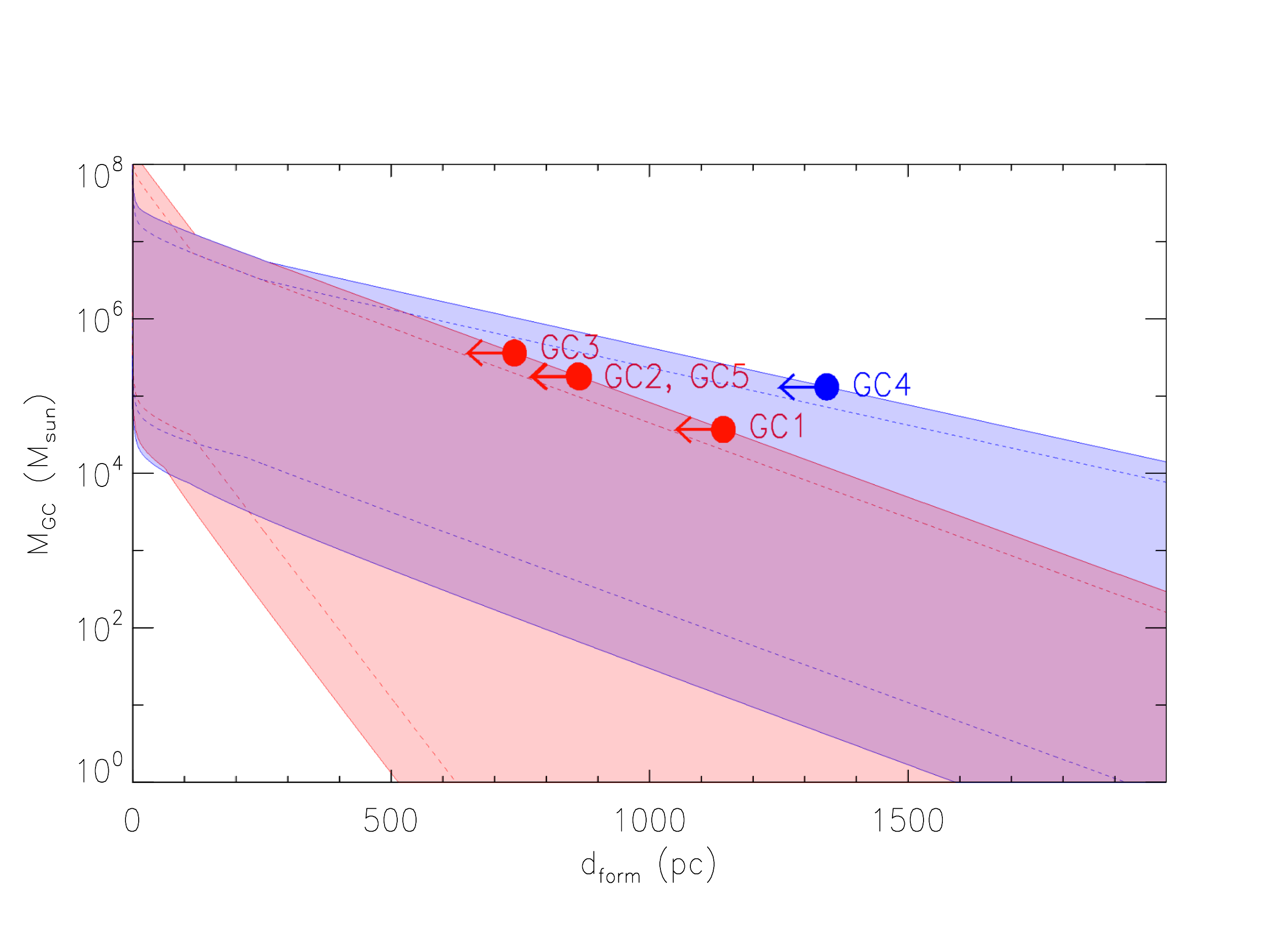}
\caption{
Hydrodynamic constraints on the formation location of the GCs in Fornax. The red and blue shaded regions show the allowed mass range of a GC to be formed at each galactic distance, at epochs representative of the formation of the $12\pm1$\,Gyr (red) and $10\pm1$\,Gyr GCs (blue). The region within the red and blue dotted lines represent the uncertainties given by the SFH and outside of them the added uncertainties from the GCs' ages.} The mass of each GC and the maximum galactic distance at which each GCs can be formed are marked with red and blue dots for the co-eval and younger GCs respectively. 
\label{fig_sfh}
\end{center}
\end{figure}

To provide physical constraints to the formation location of the GCs, we then adopt a range of possible $\tau_\mathrm{dep}$, $f_\sigma$ and $R_\mathrm{GC}$. With $0.3\,\mathrm{Gyr}<\tau_\mathrm{dep}<3\,\mathrm{Gyr}$, $0.2<f_\sigma<1.0$ and $2\,\mathrm{pc}<R_\mathrm{GC}<10\,\mathrm{pc}$ \citep{lea17}, we calculated the range of possible $M_\mathrm{GC}$ formed at different galactic radii at different time epoch. The results are shown in Figure \ref{fig_sfh}, with the red and blue region indicating the possible $M_\mathrm{GC}$ at different radii for the old GCs (GC1, GC2, GC3, GC5) and young GC4 respectively. We then consider the maximum possible formation location for each GCs, given their observed $M_\mathrm{GC}$, as the most optimistic formation distance ($d_\mathrm{form}$) from which to evolve the orbit of the GC. We derive $d_\mathrm{form}\leq$ 1144, 863, 740, 1344 and 866\, pc respectively for GC1, GC2, GC3, GC4 and GC5.

\subsection{GC Orbital Evolution}

\subsubsection{Dynamical friction implementation}\label{subsect_df}

In the seminal paper on dynamical friction by \citet{chan43}, the dynamical effect of an infalling object of mass $M_\mathrm{s}$ and velocity $v_\mathrm{s}$ through a halo of background particles moving with velocities $v_\bullet$ is described analytically as:
 
\begin{equation}
a_\mathrm{df}=\frac{d\vec{v_\mathrm{s}}}{dt} = -\frac{\pi}{2} G^{2}M_\mathrm{s}\rho_\bullet\frac{\vec{v_\mathrm{s}}}{v_\mathrm{s}^{3}}\int_0^{v_\mathrm{esc}} \frac{1}{v_\bullet}J(V) 4\pi v_\bullet^2f(v_\bullet) dv_\bullet,
\label{eq_chano}
\end{equation}
where $\rho_\bullet$ is the density of the background particles (dark matter and/or stars), $v_\mathrm{esc}$ is the escape velocity, $f(v_\bullet)$ is the velocity distribution function of the background particles and $V$ is the relative velocity between the background particle and the satellite. Due to different directions of encounter, for each $v_\bullet$, $V$ ranges from $|v_\mathrm{s}-v_\bullet|$ to $v_\mathrm{s}+v_\bullet$. $J(V)$ is an integral characterising the effect that a background particle can exert on the satellite given the different relative velocities and can be written as:
\begin{equation}
J(V)=\int^{v_s+v_\bullet}_{|v_s-v_\bullet|}\Big(1+\frac{v_s^2-v_\bullet^2}{V}\Big)\ln(1+\frac{b_{max}^2V^4}{G^2M_s^2})dV,
\label{eq_jv}
\end{equation}
where $b_{max}$ is the maximum impact parameter. Equations \ref{eq_chano} and \ref{eq_jv} can be found as Equations 25 and 26 in \citet{chan43}.

By assuming that the effect of fast-moving background particles (with $v_\bullet > v_\mathrm{s}$) is negligible, Equations \ref{eq_chano} and \ref{eq_jv} are often simplified as \citep[e.g.][]{bin87}:
\begin{equation}
a_\mathrm{df}=\frac{d\vec{v_{s}}}{dt} = -4\pi G^{2}M_{s}\rho_\bullet\ln(\Lambda)f(v_\bullet<v_{s})\frac{\vec{v_{s}}}{v_{s}^{3}},
\label{eq_chanc}
\end{equation}
where $\ln(\Lambda)$ is the Coulomb logarithm, which is given by the ratio between the maximum ($b_\mathrm{max}$) and minimum ($b_\mathrm{min}$) impact parameters as $\ln(\Lambda)=\ln(b_\mathrm{max}/b_\mathrm{min})$ and $f(v_\bullet<v_\mathrm{s})$ is the fraction of background particle that has a velocity slower the $v_\mathrm{s}$. When taking a simple assumption of the Maxwellian velocity distribution function \citep[e.g.][]{an09,petts15}, the fraction $f(v_\bullet<v_{s})$ can be expressed as:
\begin{equation}
f(v_\bullet<v_\mathrm{s})= \mathrm{erf}\Big(\frac{v_\mathrm{s}}{\sqrt{2}\sigma_\bullet}\Big)-\frac{\sqrt{2}v_\mathrm{s}}{\sqrt{\pi}\sigma_\bullet}\exp\Big(-\frac{v_\mathrm{s}^{2}}{2\sigma_\bullet^{2}}\Big),
\end{equation}
with $\sigma_\bullet$, the velocity dispersion of the background particles, being estimated by Equation \ref{eq_jeans}.
While such assumptions are generally sufficient for a cuspy dark matter profile, \citet{petts16} have recently pointed out that this is not true for cored dark matter haloes.

To show the effects of fast-moving background particles in different halo profiles, we calculated $a_\mathrm{df}$ for GC3 with $M_{GC}=3.63\times10^5\,M_\odot$ for a cuspy and a cored dark matter profile. For demonstration purpose, we adopt here the NFW profile as obtained by \citet{amo11} with phase-space modelling, with $r_\mathrm{s}=1090\,$pc and compare the derived $a_\mathrm{df}$ with a cNFW profile of the same $r_\mathrm{s}$ and $r_\mathrm{c}=1304\,$pc, which is equal to 1.75 times the stellar half-mass radius as suggested by \citet{read06}. The mass of the cNFW profile is normalised to the mass of the NFW profile at $r=r_\mathrm{c}$. These two test profiles are labelled as `\texttt{nfw0}' and `\texttt{cored0}' from hereon. We note here that there is a wide range of halo profiles derived for Fornax using various dynamical modelling technique, the \texttt{nfw0} and \texttt{cored0} profiles are merely adopted here for demonstrating the different effects of our dynamical friction treatment on cuspy and cored profiles. 

\begin{figure}
\begin{center}
\includegraphics[width=0.5\textwidth,trim=5 100 350 0,clip=true]{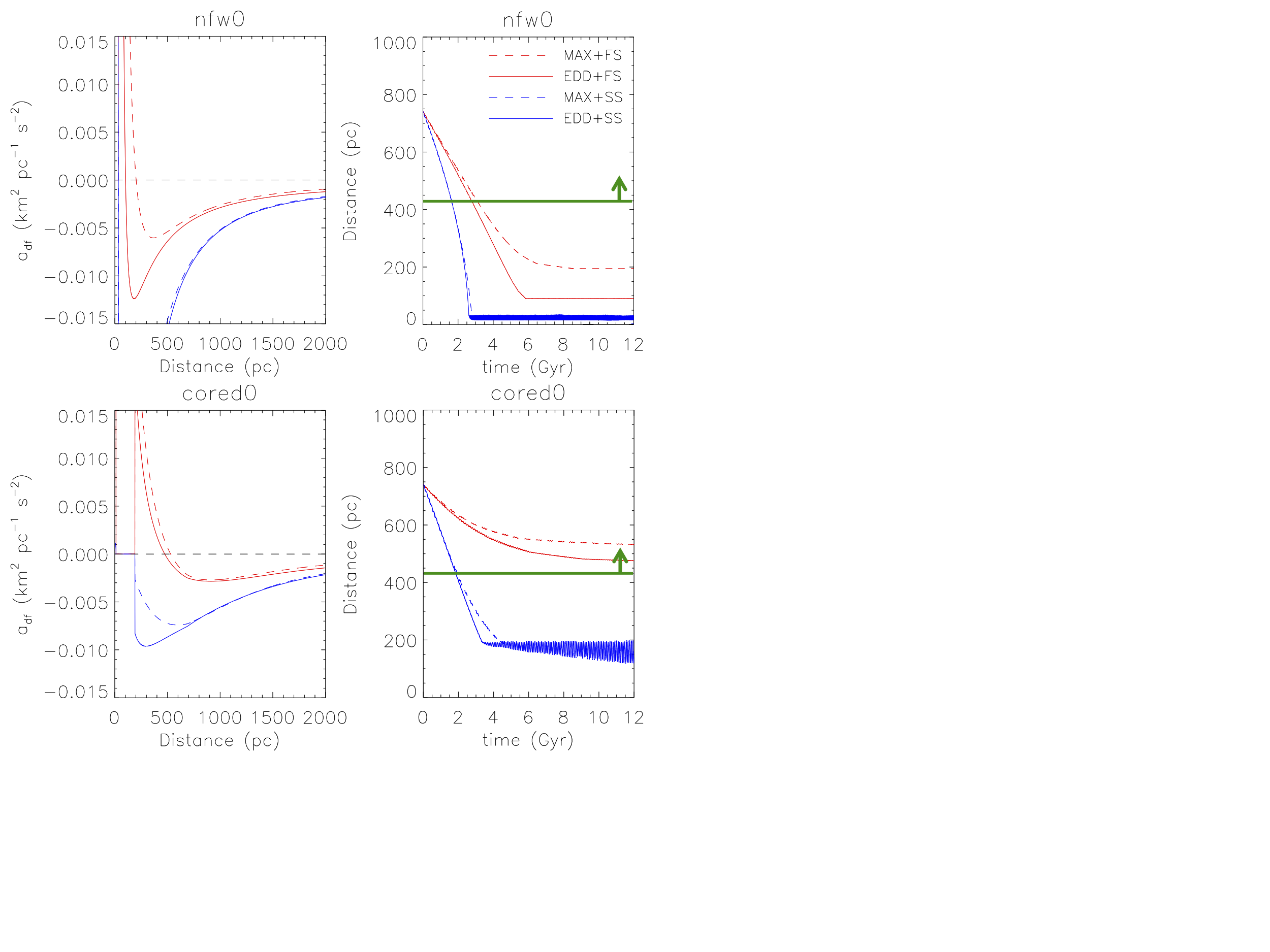}
\caption{\textit{Left:} Acceleration due to dynamical friction, $a_\mathrm{df}$, experienced by GC3 under different profile shapes and for different velocity distribution functions. Red lines denote dynamical friction treatments including fast stars (FS). Blue lines denote slow stars only (SS). Solid and dashed lines are runs using velocity distribution functions from the Eddington equation \ref{eq_edd} (EDD) and Maxwellian assumptions (MAX) respectively. \textit{Right:} the orbital decay of GC3 under the same four dynamical friction prescriptions. The green lines mark the observed galactic distance $d_\mathrm{p}$ of GC3, a lower limit of the galactocentric distance of GC3. Top and bottom row show the corresponding figures for the \texttt{nfw0} and the \texttt{cored0} profile. }
\label{fig_adf}
\end{center}
\end{figure}

The $a_\mathrm{df}$ are then derived for both profiles under the Maxwellian assumption and plotted with respect to the galactic radii in dashed lines on the left panels in Figure \ref{fig_adf}. $a_\mathrm{df}$ calculated with only the effects of slow background particles (SS) are plotted in blue and that calculated with effects from both fast and slow background particles (FS) are plotted in red. In both profiles, $a_\mathrm{df}$ is negative at large $r$, representing a dynamical friction. Where $a_\mathrm{df}=0$, we shall expect the in-spiral of the GCs due to dynamical friction to stall. Towards the inner region, $a_\mathrm{df}$ becomes positive for both haloes, albeit at very different different radii. The positive $a_\mathrm{df}$ means that when starting at these radii, the satellite will be pushed outward to where $a_\mathrm{df}=0$. The corresponding orbital decay calculated through orbital integration for GC3 starting with a circular orbit on the right panels. The orbit decays stalled at where $a_\mathrm{df} = 0$ as expected. 

We therefore see here, that the dynamical buoyancy as found by \citet{cole12} in a Fornax-like system, can be reproduced analytically by including fast-moving background particles.  While it appears at first as an exotic dynamical phenomena, it is more understandable when considering dynamical friction as a manifestation of energy equipartition, where fast-moving background particles are able to transfer kinetic energies to the infalling object. Dynamical buoyancy has a much more prominent effect in the \texttt{cored0} halo, as also found by \citet{cole12}. In contrast to the \texttt{cored0} profile, in which dynamical buoyancy exists up to $r\sim500\,\mathrm{pc}$, the dynamical buoyancy occurs at a much smaller radius of $r\sim200$\,pc in the NFW profile. This is because of the higher fraction of fast-moving particles in the inner region of the \texttt{cored0} profile. Readers interested in the derived velocity distribution function of both profiles, as well as a comparison of the analytical stalling position due to dynamical buoyancy against the results of \citet{cole12}, can refer to Appendix \ref{app_sim}.  

In addition to the stalling effects produced by the fast stars, we also include tidal stalling as shown in $N$-body simulations by \citet{in11} and described analytically by \citet{petts16}. When the GC approaches the galactocentric distance $d_{g}=r_{t}$ (where $r_{t}$ is the tidal radius of the satellite itself) the satellite will become unaffected by dynamical friction and stall. This is implemented by setting $a_\mathrm{df}=0$ when $d_{g}=r_{t}$. While tidal stalling is not important for the FS cases as $d_{g}=r_{t}$ happens within the stalling radii defined by dynamical buoyancy, it is the primary stalling mechanism for the SS cases. As pointed out by \citet{petts16}, tidal stalling is more prominent in a cored dark matter halo than a cuspy one. The same effect is seen in our models; GC3 in the SS model in the \texttt{cored0} profile stalls at $\sim$200\,pc in the \texttt{cored0} halo but $<$50\,pc in the \texttt{nfw0} halo. 

\begin{equation}
r_t^3(r_g)=\frac{GM_s}{\Omega^2-\frac{\mathrm{d}^2\Phi}{\mathrm{d}r_g^2}\big\rvert_{r_g}},
\end{equation}

\subsubsection{Velocity distribution function}\label{subsect_fv}
Both simulations and theoretical analyses have shown that dark matter haloes do not typically have a Maxwellian velocity distribution \citep[e.g.][]{ev06,han05,kuh10}. \citet{petts16} showed that such an assumption can lead to an error in $f(v_\bullet<v_{s})$ by up to $\sim$80\% depending on the halo profile. To have a more accurate handle on the velocity distribution function of various dark matter halo profiles at different radii, we therefore compute the distribution function self-consistently for an arbitrary potential by using the Eddington equation \citep{bin87}:
\begin{equation}
f(E)=\frac{1}{\sqrt{8}\pi^{2}}\int_0^E\frac{d^{2}\rho}{d\Phi^{2}}\frac{d\Phi}{\sqrt{E-\Phi}},
\label{eq_edd}
\end{equation}
where $E$ is the relative energy, $E=\Phi-mv^{2}/2$.
We show $a_\mathrm{df}$ calculated for GC3 in the \texttt{nfw0} and \texttt{cored0} profiles with the the Eddington velocity distribution function in solid lines. Again, the blue curve shows the results obtained for the SS case and the red curve shows the results obtained for FS case. Notice here that adopting a Maxwellian assumption will lead to an overestimation of the stalling radius by a factor of 2 in the NFW halo. For the rest of the paper, we shall adopt the improved dynamical friction treatment using velocity distribution function calculated from the Eddington equation and taking into account the effects from fast stars (i.e. case EDD+FS).

\subsubsection{Orbit integration}
Starting at a galactocentric distance in Fornax determined as described in Section \ref{subsect_rstart}, we then integrate the orbit of each GC, subjecting to dynamical friction/buoyancy as described in Section \ref{subsect_df}, as well as the gravitational acceleration given by the underlying potential of Fornax. The orbit integration continues for the respective ages of each of the GCs. The positions, velocities and accelerations are updated at every time step of 1\,kyr, with a precision of 0.01\,pc and 0.01\,km\,s$^{-1}$. We assume circular orbits for the GCs to study the most conservative case as GCs on more eccentric orbits would simply be subjected to a severer dynamical friction. The numerical integrations are done with the Python module \texttt{odeint} from \texttt{scipy}. 

\subsection{A past merger event}\label{subsect_merge}
\begin{figure*}
\begin{center}
\includegraphics[width=0.95\textwidth,trim=0 300 20 20,clip=true]{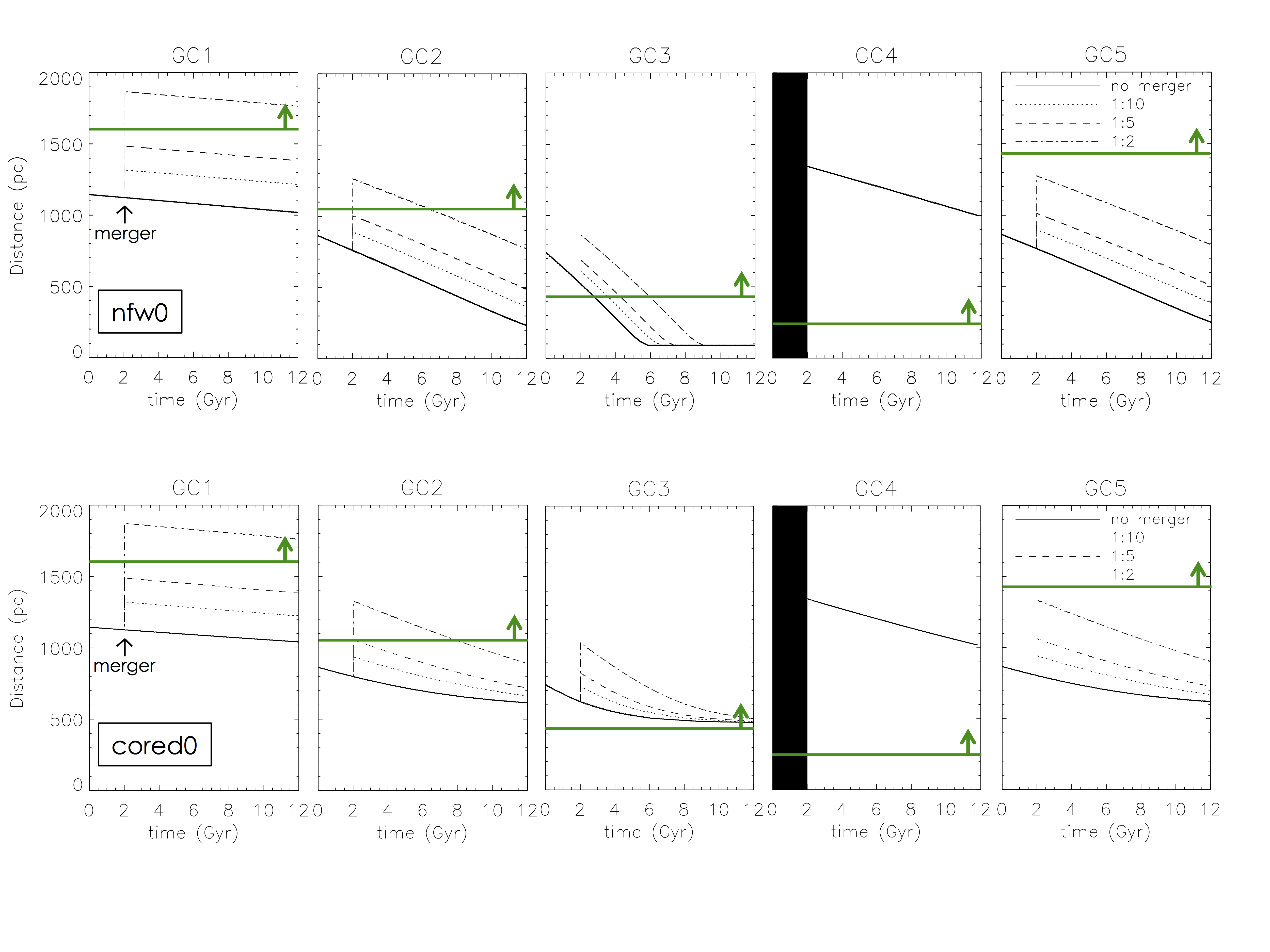}
\caption{Orbital evolution of GCs in the \texttt{nfw0} halo with various merger histories. The green lines mark the observed present day distance $d_\mathrm{p}$ of each GC, a lower limit of the their galactocentric distances.Under this profile, only GC4 can survive outside of its observed distance without a merger, GC1 would need an 1:5 merger and GC2 an 1:2 merger. Both GC3 and GC5 would need a merger with an even more substantial mass ratio than 1:2 to exist outside of its $d_\mathrm{p}$ under this profile.
\label{fig_result_nfw}}
\end{center}
\end{figure*}

\begin{figure*}
\begin{center}
\includegraphics[width=0.95\textwidth,trim=0 50 20 270,clip=true]{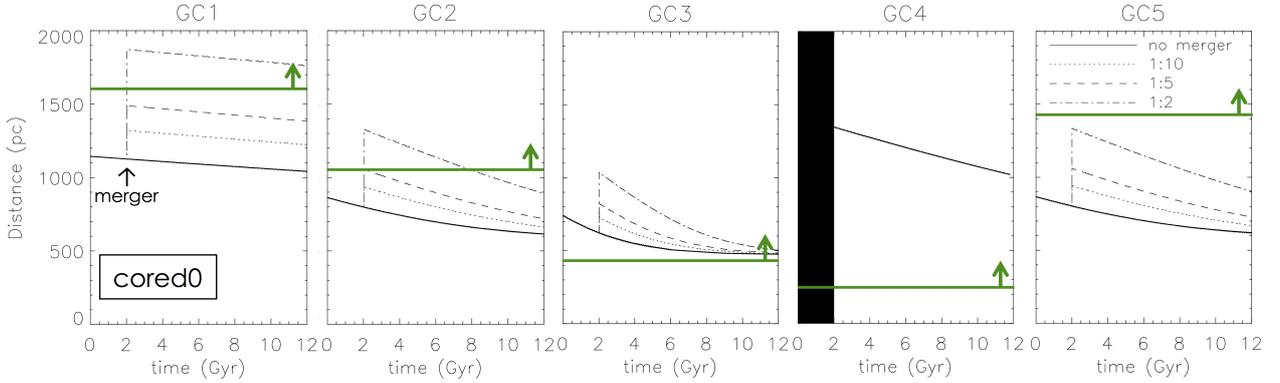}
\caption{Same as Figure \ref{fig_result_nfw}, but for the \texttt{cored0} halo profile. Under this profile, both GC3 and GC4 can survive outside of its observed distance without a merger. As with the NFW profile, GC1 would need an 1:5 merger and GC2 an 1:2 merger. GC5 would however still need a merger with a mass ratio smaller than 1:2 to exist outside of its $d_\mathrm{p}$ under this profile.
\label{fig_result_cored}}
\end{center}
\end{figure*}

The complex stellar morphology, metallicity and age distribution of Fornax suggests the galaxy might have experienced a significant merger event. A dry merger can significantly expand the final system size, given that the dominant stellar and dark matter components are non-dissipative. This could cause the GC (and stellar and DM) orbits to expand. In other words, a non-dissipative merger would have allowed the GCs to acquire a larger present-day galactic distance than the pressure equilibrium criteria allow for. To estimate this effect, we adopt the analytic expansion derived in \cite{naab09}:
\begin{equation}\label{eq_mr}
\frac{r_\mathrm{f}}{r_\mathrm{i}}=\frac{(1+\eta)^{2}}{(1+\eta\epsilon)},
\end{equation}
where $r_\mathrm{i}$ and $r_\mathrm{f}$ are the position of the GCs before and after the merger, $\eta$ and $\epsilon$ are determined by the merger ratio with:
\begin{equation}
\eta = \frac{M_\mathrm{acc}}{M_\mathrm{host}}; \epsilon = \frac{\langle v_\mathrm{acc}^{2}\rangle}{\langle v_\mathrm{host}^{2}\rangle},
\end{equation}
where $M_\mathrm{acc}$, $\langle v_\mathrm{acc}^{2}\rangle$, $M_\mathrm{host}$ and $\langle v_\mathrm{host}^{2}\rangle$ are the mass and velocity dispersion of the accreted and host galaxies. We have assumed the Faber-Jackson relation of $M\propto \sigma^4$ \citep{fab76} when calculating $\langle v_\mathrm{acc}^{2}\rangle$ and $\langle v_\mathrm{host}^{2}\rangle$.\footnote{It has been suggested that the index of the Faber-Jackson relation $\alpha$ in low-mass galaxies can be as low as $\sim$2 \citep[e.g.][]{kou12}. The differences in $r_\mathrm{f}/r_\mathrm{i}$ between $\alpha=4$ and $\alpha=2$ can be written as $(1+\eta^{1.5})/(1+\eta^2)$. Within our tested range of $0.0<\eta<0.5$, it amounts to an 8\% change in the final to initial position ratio.}

Given the younger age and significantly higher metallicity of GC4 \citep{deB16}, we consider the case in which a dry merger happened 10\,Gyrs ago which triggered the formation of GC4 from small amounts of residual gas in the total system\footnote{Naively assuming a star formation efficiency per free-fall time for a molecular cloud of $\epsilon_{ff}=0.03$, this would require $M_{gas} \geq 5\times10^{6}\,M_\odot$ in Fornax at the time.  This is reasonable given that Fornax continued to form another $5\times10^{6}\,M_\odot$ of field stars at a low level for another $\sim$9\,Gyrs after this and so clearly retained some gas.}. GC1, GC2, GC3 and GC5 will hence experience an orbital expansion due to the merger while GC4 will not. We demonstrate how our simple analytic expansion from Equation \ref{eq_mr} would affect the final GCs positions in Figure \ref{fig_result_nfw} and \ref{fig_result_cored} for the \texttt{nfw0} and \texttt{cored0} profiles respectively. We tested four different scenarios: no merger (solid lines), an 1:10 merger (dotted lines), an 1:5 merger (dashed lines) and an 1:2 merger (dash-dotted lines). GC1, GC2, GC3 and GC5 would hence for the first 2\,Gyrs orbit through a dark matter halo with a viral mass $1-\eta$ times the current day virial mass, the $r_\mathrm{s}$ and $r_\mathrm{c}$ of the dark matter profile before the merger also scale as Equation \ref{eq_mr}. As shown in Figure \ref{fig_result_nfw} and \ref{fig_result_cored}, the expansion experienced by the GCs increases as the mass ratio between the host and accreted galaxy decreases. By comparing the modelled present location of the GCs with the observed projected presented-day position $d_\mathrm{p}$ (horizontal dashed line), the \texttt{nfw0} profile can be ruled out because both GC3 and GC5 end up inside their respective $d_\mathrm{p}$s even with an 1:2 merger. As for the \texttt{cored0} profile, GC3 can survive out of its $d_\mathrm{p}$ but GC5 still fails to do so even with an 1:2 merger.

\section{Results}\label{sect_results}
In this section we will show the results of the orbital evolution for the GCs in Fornax. We run our semi-analytic model on a grid of dark matter halo profiles with $r_\mathrm{s}$ and $r_\mathrm{c}$ each drawn from 1000$-$6000\,pc in steps of 1000\,pc. For each halo profile we include a `no merger' case and three merger cases with merger mass ratios of 1:10, 1:5 and 1:2. We then compare the modelled present day galactic distance of each GCs with the observed $d_\mathrm{p}$. 

\begin{figure*}
\begin{center}
\includegraphics[width=1.05\textwidth,trim=0 375 10 0,clip=true]{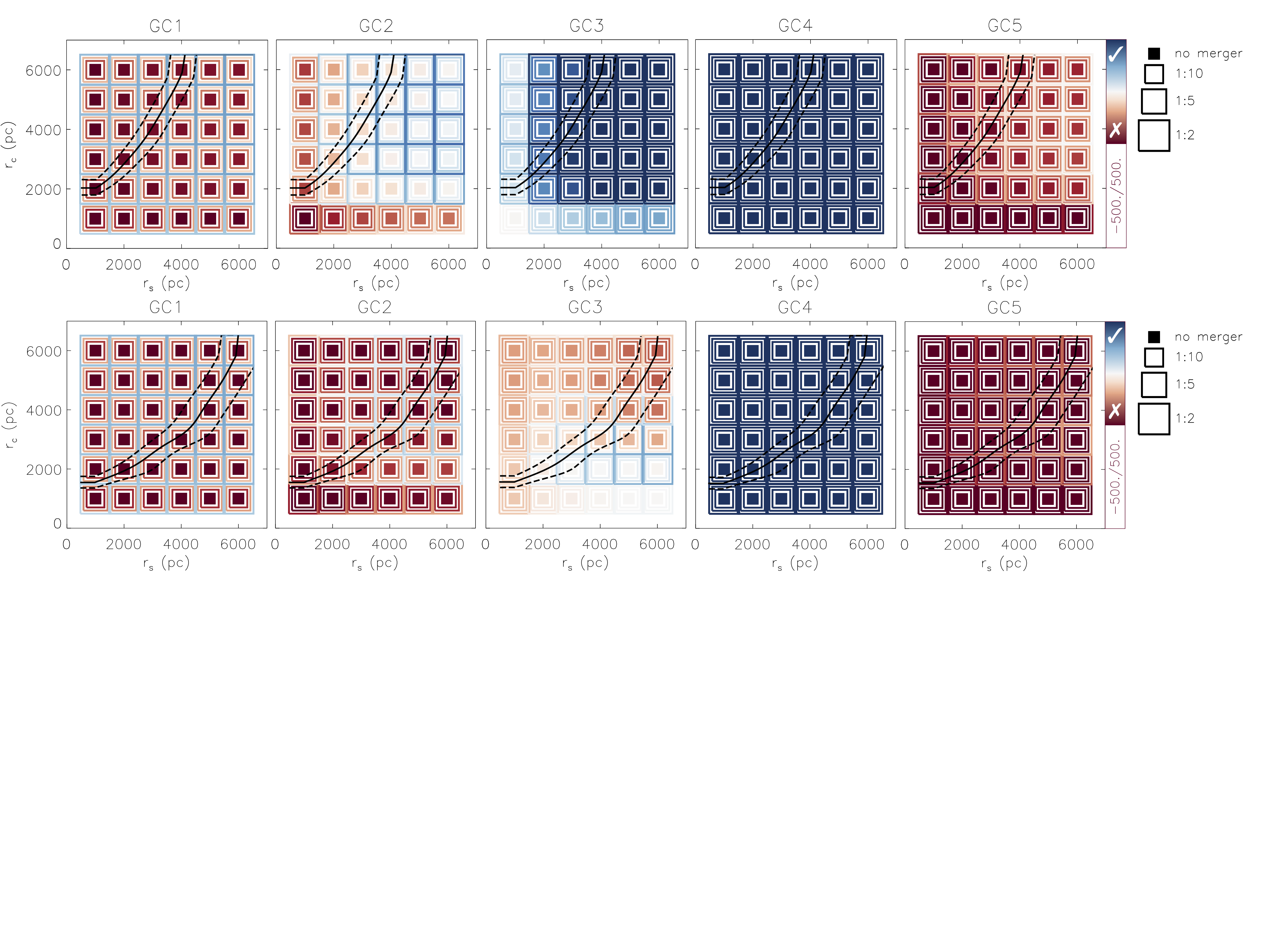}
\caption{Summary of orbit integration for the grid of DM halos and merger mass ratios.  Each plot is for a different GC and shows the grid of DM halo scale radii $r_\mathrm{s}$ and core radii $r_\mathrm{c}$ for that trial. For each ($r_\mathrm{s},r_\mathrm{c}$) pair, we have run the dynamical friction model under the assumption that Fornax has experienced no merger (filled squares), a 1:10 merger, a 1:5 merger, and a 1:2 merger, with the merger mass ratio indicated by the size of the square. In each trial, the final position oof the GC relative to its observed present day is indicated by the colour of the box. The models marked with blue means that the GC is found to survive outside the $d_\mathrm{p}$ (as marked by tick on the colour bar) and hence suggesting the particular parameters ($r_\mathrm{s}, r_\mathrm{c}, \eta$) represent a plausible dark matter profile and merger history for Fornax. The halo parameters which follow an $M_\mathrm{vir}$-concentration relation inferred from cosmological simulations \citep{dut14} are shown in the background marked with black contours, the dashed contours mark the 5-$\sigma$ values.
\label{fig_grid}}
\end{center}
\end{figure*}

The results are presented in Figure \ref{fig_grid}. The size of the squares represents the merger ratio. The colour coding represents the difference between the final model galactocentric distance and the current projected distance, $d_\mathrm{p}$ for each GCs. Blue implies that the modelled distance is outside of the observed $d_\mathrm{p}$, meaning that the dark matter halo with parameters ($r_\mathrm{s}, r_\mathrm{c}$) is plausible given the corresponding merger with mass ratio $\eta$ had happened. 

The observed galactocentric distance of both GC3 and GC4 can be well reproduced with any of the dark matter halo profile, without the need for any merger. The addition of a past merger event does not significantly change the required ($r_\mathrm{s}, r_\mathrm{c}$) for GC3. This is because of the large mass of GC3, which implies that the dynamical friction timescale is relatively short compared to the other GCs. Therefore, GC3 reaches its stalling radius within a Hubble time regardless of the merger. This is also reflected in Figure \ref{fig_result_cored}, which shows that the final position of GC3 under different merger ratios all converge to the stalling radius of the \texttt{cored0} profile. It is different for GC1, GC2 and GC5 because their masses are a half to an order of magnitude smaller than GC3, allowing them to have a much longer dynamical timescale. The orbital expansion given by the merger event therefore has more importance on these final GC positions. 

\begin{figure}
\begin{center}
\includegraphics[width=0.55\textwidth,trim= 185 20 150 190,clip=true]{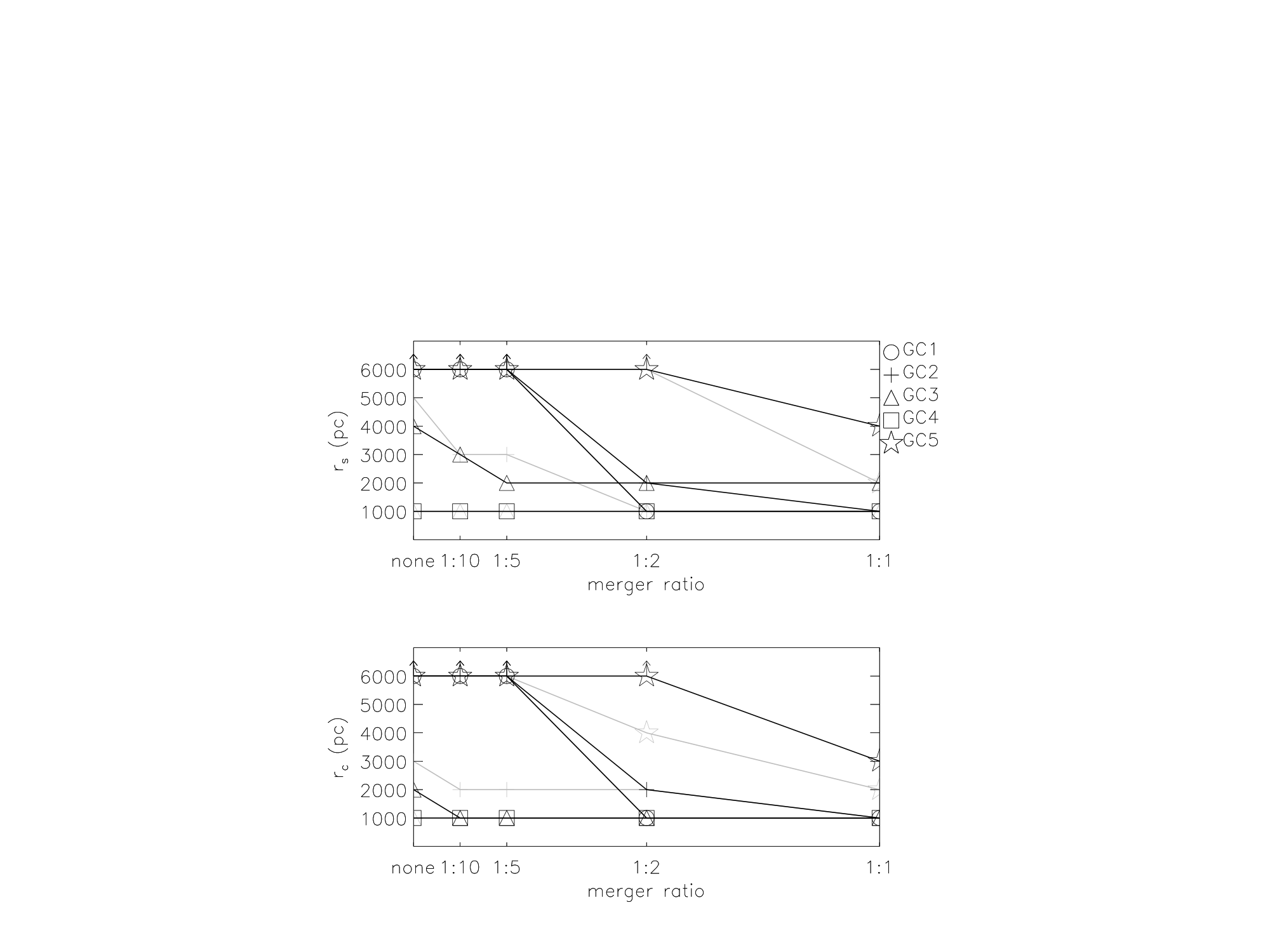}
\caption{Minimum $r_\mathrm{s}$ (top panel) and $r_\mathrm{c}$ (bottom panel) of Fornax dark matter halo as constrained by the five GCs under the various merger scenarios. The grey lines correspond to the DM only case while the black lines correspond to the DM+stars case.}
\label{fig_res}
\end{center}
\end{figure}

Given a merger with mass ratio 1:2, the observed $d_\mathrm{p}$ of GC1 can also be reproduced with any of the dark matter profiles. Without that, none of the tested profile can reproduce the observed $d_\mathrm{p}$ for GC1. The minimum ($r_\mathrm{s}, r_\mathrm{c}$) required for GC2 is (5000, 3000)\,pc in the 'no merger' case, (3000, 2000)\,pc with an 1:10 or an 1:5 merger and (3000, 1000)\,pc for a 1:2 merger. Finally, the observed $d_\mathrm{p}$ of GC5 can only be reproduced with a merger of mass ratio 1:2 at $(r_\mathrm{s}, r_\mathrm{c})>(6000, 4000)$\,pc. The minimum $r_\mathrm{s}$ and $r_\mathrm{c}$ as required by each GC is plotted in Figure \ref{fig_res}, with the case for a 1:1 merger marked as an additional reference in this plot.

\begin{figure}
\begin{center}
\includegraphics[width=0.47\textwidth,trim=368 10 58 25,clip=True]{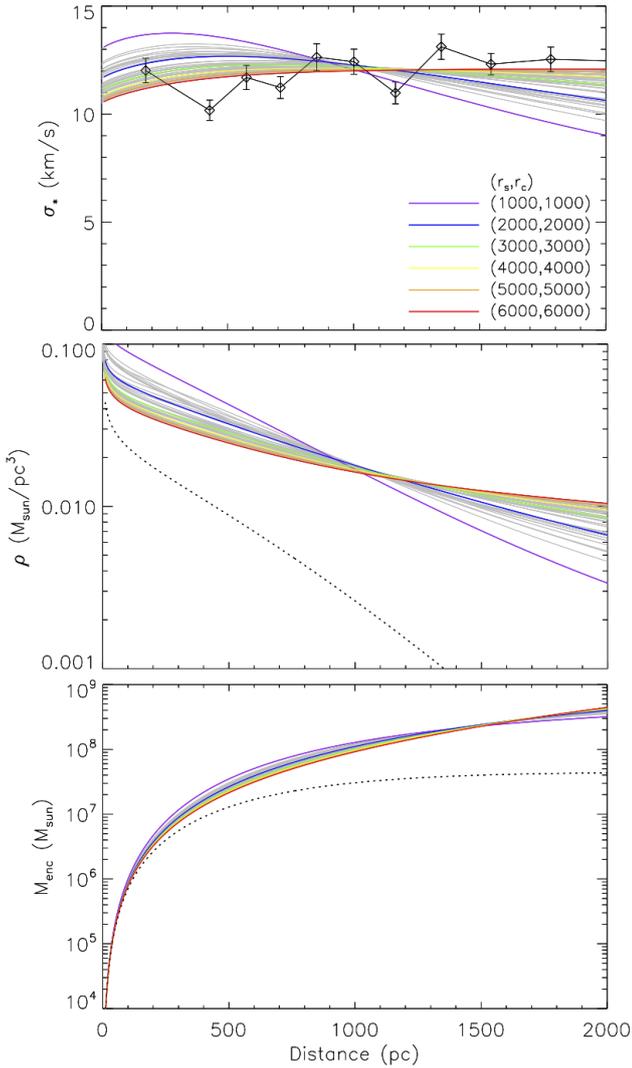}
\caption{Top: the observed $\sigma_\star$ of Fornax is plotted in black diamonds with error bars. Overlaid in grey are all the dark matter profiles we tested of our ($r_\mathrm{s} ,r_\mathrm{c}$) grid, for which we added a stellar component. We show in colour are six examples of the $\sigma_\star$ profiles from our normalised dark matter profiles. Middle and bottom: the corresponding density and enclosed mass profiles. The dotted black lines in the middle and bottom panel show the stellar density and enclosed mass profiles respectively.}
\label{fig_sigma_sd}
\end{center}
\end{figure}

\begin{figure*}
\begin{center}
\includegraphics[width=1.05\textwidth,trim=0 210 10 170,clip=true]{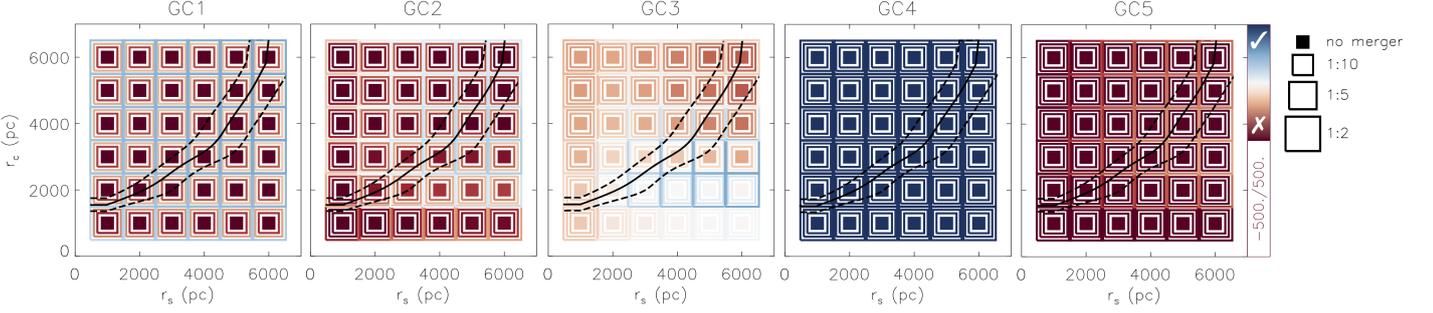}
\caption{Same as Figure \ref{fig_grid}, but with a stellar component included in the background mass profile. 
\label{fig_grid_sd}}
\end{center}
\end{figure*}

While these results are run with only the dark matter halo contributing to the potential, the stellar contribution within the tidal radius of Fornax is expected to be non-negligible. Therefore we repeat the exercise and include the stellar component as described in Section \ref{subsect_bg}. The dark matter haloes of each ($r_\mathrm{s}, r_\mathrm{c}$) are renormalised with the inclusion of the stellar component using the observed $\sigma_\star(R)$ as described also in Section \ref{subsect_bg}. We show in Figure \ref{fig_sigma_sd} the normalisation of $\sigma_\star(R)$ with the inclusion of a stellar component, and the corresponding density and enclosed mass profiles of our $(r_\mathrm{s}, r_\mathrm{c})$ grid. The resultant differences between the final model galactocentric distances and $d_\mathrm{p}$s are shown in Figure \ref{fig_grid_sd}. In general, either a larger $(r_\mathrm{s}, r_\mathrm{c})$ or a smaller mass ratio in the merger is required due to the fact that the stellar component tends to steepen the overall density profile. This is true in particular for GC2, GC3 and GC5, where the profile shape has a noticeable effect on the final location of the GC. GC4 is still permitted under all halo profiles, due to its small present-day galactocentric distance. As for GC1, the small GC mass leads to a long dynamical friction timescale, which means that the merger ratio has a more prominent effect on the final GC location than the underlying density profile. As in the case of DM only, GC1 requires a merger ratio of 1:2 to allow the final location of the GC to be outside of the present-day observed distance $d_\mathrm{p}$. GC2 requires at least an 1:2 merger, with which a $(r_\mathrm{s}, r_\mathrm{c})$ of (2000, 2000)\,pc is sufficient for the GC to end up outside of $d_\mathrm{p}$. GC3 requires a minimum merger mass ratio of 1:10 at $(r_\mathrm{s}, r_\mathrm{c})$ of (3000, 2000)\,pc, a mass ratio of 1:2 allows a ($r_\mathrm{s}, r_\mathrm{c}$) of as small as (2000, 1000)\,pc. GC5 now becomes problematic under all halo profiles and they are not permitted to exist outside of $d_\mathrm{p}$ with any merger with mass ratios larger than 1:2.

\section{Discussion}\label{sect_diss}
With constraints from the $d_\mathrm{p}$ of the GCs in Fornax, our semi-analytic orbital evolution model suggests that there is a dark matter core of size no smaller than 1000\,pc in Fornax, and that the galaxy has experienced a past merger of mass ratio more substantial than 1:5. In this section, we first present a self-consistent picture for the co-evolution of Fornax and its GCs in Section \ref{subsect_pic}. We then provide additional evidences from the chemistry of Fornax to support the merger scenario in Section \ref{subsect_mdf} and our proposed origins of the GCs in Section \ref{subsect_mgcfs}. Section \ref{subsect_mged} concerns with evidences of dwarf-dwarf mergers both from cosmological simulations and observed interactions between dwarfs. In Section \ref{subsect_ndm} we compare our derived dark matter halo profile with cosmological simulation results and discuss the implications of the apparent large dark matter core on the nature of dark matter. We close this section by presenting some caveats of this work.

\subsection{A self-consistent picture for the co-evolution of Fornax and its GCs}\label{subsect_pic}
GC5 stands out as the only GC that would require a ($r_\mathrm{s}, r_\mathrm{c}$) larger than our explored range of value. The younger age and higher metallicity of GC5 when compared with GC1, GC2 and GC3 also might be hinting at a different origin of this GC (Section \ref{subsect_mdf}). We propose the following scenario for the co-evolution of Fornax and its GCs: (1) GC1, GC2 and GC3 were formed in a proto-Fornax at $\sim$12\,Gyrs ago, (2) GC5 was formed $\sim$11\,Gyrs ago in the lower mass dwarf galaxy that will go on to merge with the proto-Fornax, and (3) the merger which happened $\sim$10\,Gyrs ago triggered the formation of GC4, and at the same time deposits GC5, and scatters GC1,2,3 to larger orbits conducive to their survival. 

The existence of a sixth GC has recently been re-discussed by \citet{wang19}, where they show with deep DECam imaging data that a past association of stars is likely to be a star cluster with stellar mass of $M_{*} \sim 10^{4}\,M_{\odot}$. This object has a projected distance of $d_{p}$ of 270\,pc and its metallicity is inferred through photometry to be similar to GC4 ([Fe/H]$\sim -1.4$). Notably, its low mass but small projected distance is at odds with naive expectations for dynamical friction (especially relative to the higher mass, but further out GCs).  While further work on the orbit and ages of this GC will be necessary to fully understand its role in the evolution of Fornax, we note that its central position and relatively high metallicity (compared to other GCs) can be naturally explained with our merger scenario: just like GC4, GC6 would be a product of triggered star formation due to compression of gas in the dwarf-dwarf merger approximately 10\,Gyrs ago and reside close to the center of Fornax after that event.

\subsection{The stellar mass in GCs and field stars in the context of a past merger}\label{subsect_mgcfs}

In addition to a surprisingly large number of GCs, Fornax notably shows a very high fraction of mass in star clusters relative to low metallicity field stars \citep{lar12}.  This provides strong constraints on the amount of mass loss and initial mass of GCs, which is of extreme importance for theoretical explanations of the multiple population phenomena in GCs \citep[c.f.,][]{bas17}.

The top panel of Figure 14 shows the cumulative mass in Fornax's five GCs relative to its field stars as a function of cumulative metallicity ($M_\mathrm{GC}/M_\mathrm{\star, gal}<$[Fe/H]).  Here we have used the observed SFH of Fornax (corrected for spatial completeness) and the observed mean Age-Metallicity relation (AMR) of the field stars \citep{bat06, lea13} to compute $M_\star$ as function of [Fe/H].  We plot this versus the age of the stellar populations (from the field star AMR) and the age of the GCs from isochrone fitting \citep{deB16}. \cite{lar12} computed the mass fraction based in GCs relative to field stars by analysing the MDF of the field stars directly and making corrections for sample selection and stellar evolution effects.  Here we find comparable qualitative results when using the stellar mass growth for the galaxy itself derived from the SFH of Fornax and the spectroscopic AMR.

The magenta line shows the values for a galaxy with the observed SFH and chemical enrichment which formed a single GC of $M_{GC} = 2\times10^{5}$ at any point in time.  The offset blue lines show what values would be expected if you formed the same mass GC in a dwarf galaxy that was some stellar mass ratio $1:10 \leq \eta_{*} \leq 1:1$ less than Fornax.  This is computed by simply shifting the AMR by an amount based on the observed Local Group mass metallicity relation \citep[e.g.][]{kir13}. This toy exercise suggests that a dwarf with stellar mass $\sim 1/3$ of that of Fornax and a single GC would have values similar to where GC5 sits on this diagram.

\begin{figure}
\begin{center}
\includegraphics[width=0.45\textwidth,trim= 0 0 0 0,clip=true]{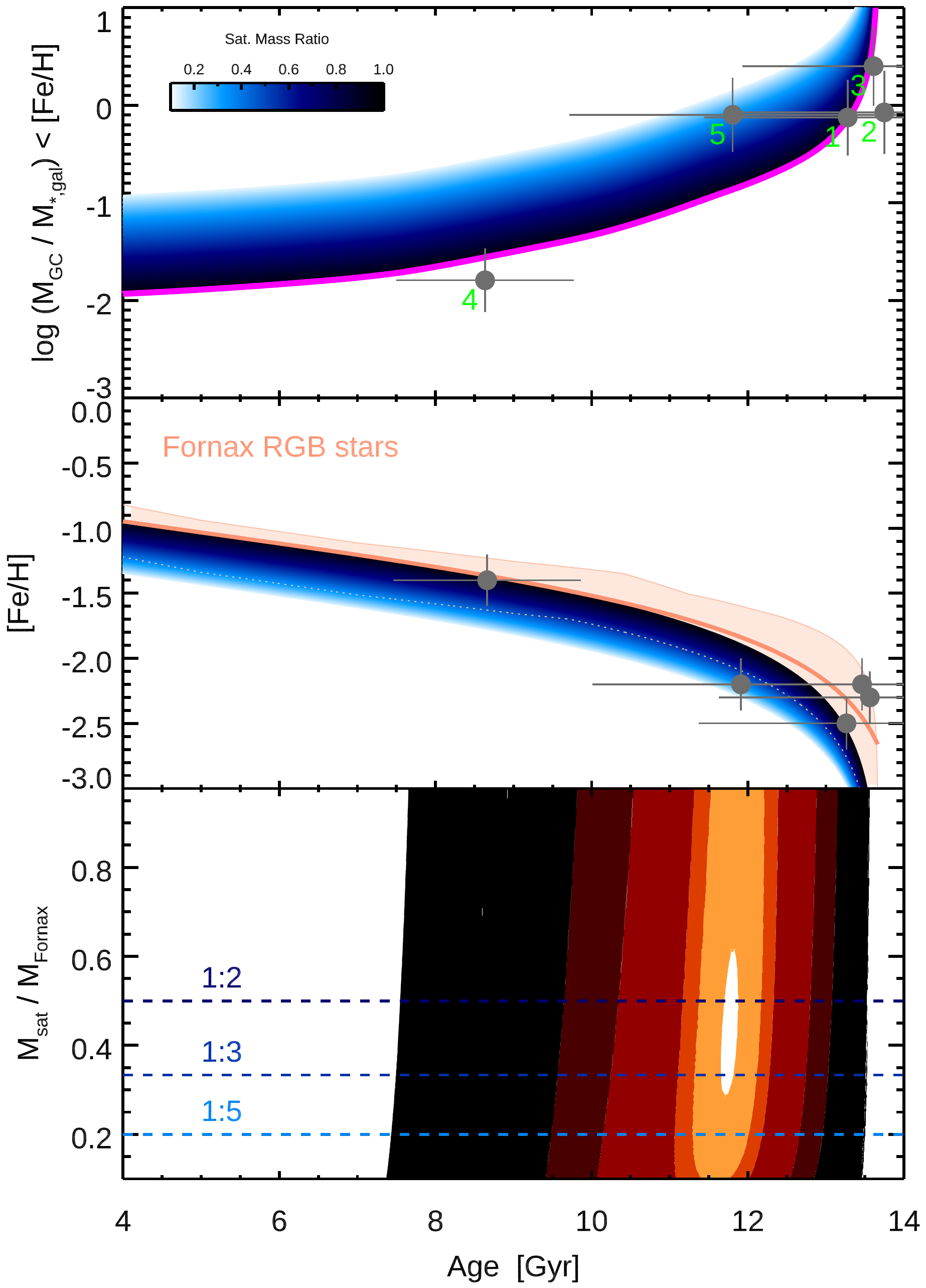}
\caption{\textit{Top:} the cumulative mass in GC stars relative to Fornax fields stars below that [Fe/H] value plotted in magenta, with grey dots showing this ratio within each GCs; Middle: observed age-metallicity relation of the RGB field stars in Fornax (orange). In the top and middle panel, the blue band represents the change in the plotted quantities for a dwarf of mass ratio $\eta$, ranging from 0 (light blue) to 1 (dark blue), which had a single GC. \textit{Bottom:} The offset location of GC 5 in the top and middle panels suggests a required mass ratio close to the one derived from our dynamical model. The contours are in fractions of [0.1, 0.5, 0.7, 0.9,0.95,0.99] of the maximum likelihood.
\label{fig_gcmerge}}
\end{center}
\end{figure}

Another way to compare the GC and field stellar population is by looking at their AMRs. In the middle panel of Figure \ref{fig_gcmerge} we show the observed AMR of the Fornax RGB stars as the orange band. The observed AMR closely follows a leaky box analytic chemical evolution model, and similar to the top panel, we show in blue the implied AMR for dwarf galaxies of smaller total stellar masses using the same shifted empirical mass-metallicity relations. As above, the corresponding observed ages and metallicities for the GCs are plotted in grey dots. Once again GC5 is an outlier with respect to the field stars' AMR, and corresponds more closely to the chemistry of a dwarf galaxy of mass $\sim$1/3 of Fornax. 

The bottom panel shows a summary of the implied mass ratios which are more chemically consistent with GC5. Merger ratios of 1:2, 1:3 and 1:5 are marked by magenta dashed lines. The analysis here suggests that not only would a merger of mass ratio 1:2 to 1:5 allow GC5 to survive outside of it observed projected distance, but that it also is consistent with the mass and stellar populations in GC5 and those of the field stars in Fornax.

In the scenario we present (GC 5 accreted, GC 4 formed from in-situ gas later), the mass of GCs 1,2,3 are still more than 10\% of the metal poor population of Fornax.  This scenario then does not fully alleviate the high GC-field star mass ratio found by \cite{lar12}. In addition, in our merger scenario the lower mass satellite which merged with Fornax was more metal poor (See Appendix \ref{subsect_mdf}) and so should contribute ex-situ field stars to the lowest metallicities.  This would increase the mass of GCs 1,2 and 3 relative to the Fornax field population.  For our two toy-examples of mergers in Figures \ref{fig_carwlm} and \ref{fig_sclwlm}, the mass ratio between GCs 1,2, and 3 and the in-situ low metallicity of the proto-Fornax which hosted them, would still be over 20\% - consistent with the findings and constraints on GC mass loss identified by \cite{lar12}.

\subsection{Additional evidence of dwarf-dwarf mergers}\label{subsect_mged}

It has been shown in cosmological zoom-in simulations that group processing such as mergers of gas-rich dwarf irregulars (dIrrs) is a possible formation pathway for gas-poor dSphs like Fornax \citep[e.g.][]{wet15}. With cosmological simulations, \cite{ben16} show that such a process may explain the overall metallicity gradients found in some dSphs that are caused by the different spatial distribution of the multiple stellar populations; where young and concentrated metal-rich components are surrounded by older and metal-poorer stars, as seen in for example Sextans \citep{tol04, bat11}, Sculptor \citep{bat08}, as well as Fornax \citep{bat06}. In the merging process, the older and more metal-poor stars can be dispersed, leading to a larger spatial distribution and lower central density, as compared with the younger metal-rich population formed after the merger. Hence, in some cases dwarf-dwarf mergers may be a channel to produce stellar population gradients in low mass galaxies.

Observational evidence of dwarf-dwarf mergers is also becoming increasingly common. The TiNy Titans Survey (TNT) found evidence of interactions between isolated pairs of dwarf galaxies, such as disturbed optical and HI morphologies, as well as images of dwarf pairs on the verge of merging \citep{stier15}. The Magellanic Clouds have been shown to host a rich satellite system in recent surveys, such as the Dark Energy Survey \citep[DES;][]{bech15, kop15} and the Survey of the MAgellanic Stellar History \citep[SMASH;][]{mar15}. These works suggest that dwarf galaxies can have satellites of their own that may later be assimilated. \cite{amo14} kinematically detected a stellar stream in the dSph Andromeda II (And II) of which the progenitor is possibly a dwarf galaxy with similar mass as And II, indicating a past major merger. Differential rotation between the metal-rich and metal-poor stars in the dSph Sculptor is also possibly a result of a past merger \citep{zhu16}. \cite{cic18} also found merger evidences in the dSph Sextans, where a ring-like stellar feature shows higher-than-average line-of-sight velocities and lower-than-average metallicities, while \cite{kach17} found evidence of prolate rotation in the Phoenix dSph.

Specifically to Fornax, \cite{amo12} (AE12 hereafter) suggest signatures of three stellar populations from its complex MDF, and show that there is a 40$\degree$ difference in the rotation axes between the metal-poor (MP) and the intermediate-metallicity (IM) populations which imply counter-rotation. The authors have attributed such complexities to a merger of a bound pair, with the companion, represented by the MP population, comprising a fraction of $0.31\pm0.06$ of the spectroscopic sample of stars. Given the uncertainties on the complete spectroscopic selection function for Fornax, to compare to our work we bound the possible mass fraction of this population by: 1) multiplying this fraction directly with the total stellar mass of Fornax (likely an upper limit), or by 2) following AE12 and multiplying the observed luminosity of the RGBs in the MP population by 6\footnote{AE12 assume that 1/3 of the MP giants reside in the metal poorest tail and that the RGB luminosities is 1/2 of the total luminosity.} and then applying a mass-to-light ratio $M_\star/L$ of 2 \citep{mc12}. This analysis yields a stellar mass of $3\times10^{6}-1.5\times10^{7}\,M_\odot$ for the MP population, which could comprise the lower mass merging fragment. Given the observed age-metallicity relation for Fornax, the pre-merger proto-Fornax is plausibly represented by the IM population, which comprises a fraction of $0.56\pm0.05$ of the spectroscopic sample. A similar computation for this population results in a proto-Fornax stellar mass $1\times10^{7}-2.5\times10^{7}\,M_\odot$. 
As a qualitative example, Sculptor and WLM, with stellar masses of $7\times10^{6}\,M_\odot$ \citep{ber18} and $0.9-1.8\times10^{7}\,M_\odot$ \citep{lea17} respectively at $z\sim2$, fall right into the ranges suggested by the chemodynamical analysis of AE12. Their combined metallicity distribution function can also reproduce the shape of that of Fornax (see Appendix \ref{subsect_mdf}) and therefore could be considered as potential analogues to the companion and proto-Fornax respectively. With a dynamical mass ratio of $\sim$1:3 at $z\sim2$ \citep[][and the references therein]{lea17, ber18}, the merger mass ratio of 1:2 to 1:5 inferred from our dynamical friction analysis is therefore consistent with the results from AE12. Given that \cite{bat06} can associate most of the more metal poor component with an old age of $>$10\,Gyrs, it is therefore plausible that the merger fragment stopped forming stars at $\sim$10\,Gyrs ago, indicating an early merger around that time for Fornax. While there is additional evidence for shell-like substructures in the central region of Fornax \citep{col05}, the young ages and high metallicities of these features and their significant pre-enrichment \citep{col08}, as pointed out by \cite{amo12}, suggest that the features are formed from self-enriched gas of Fornax itself at late times, rather than due to an accretion event. Furthermore, they suggested that it is unlikely to have two sub-haloes colliding with one another in the Milky Way halo at such a late time when the collision energetics is considered.

\subsection{Implication for the nature of dark matter from the derived halo profile}\label{subsect_ndm}
The conditions for GC survival in Fornax require a particular form of the dark matter halo. Here we briefly discuss how this may place constraints on the self-interacting nature of dark matter. To provide a comparison of the required $(r_\mathrm{s}, r_\mathrm{c})$ with respect to dark matter halo parameters in $\Lambda$CDM cosmological simulations, we show the mass-concentration ($M-c$) relation as seen in such simulations as a black contour in Figure \ref{fig_grid} and \ref{fig_grid_sd}. We adopt here the $M-c$ relation from \citet{dut14}. The concentration of our dark matter haloes are calculated as $c=r_\mathrm{200}/r_{-2}$, where $r_{-2}$ is the radius at which the logarithmic slope of the density profile equals -2. With a merger mass ratio of 1:2, GC1, GC2, GC3 and GC4 can all survive outside their respective $d_\mathrm{p}$ with a dark matter profile that lies on the $M-c$ relation, of ($r_\mathrm{s}, r_\mathrm{c}$)$\sim$(2000, 2000)\,pc. 

We next check whether the required core size is compatible with dark matter cores created by baryonic feedback processes, such as those seen in $\Lambda$CDM hydrodynamical simulations of dwarf galaxies. As \cite{read16} have shown, the dark matter core size in their simulations is approximately 1.75 times of the half-light radius. In the case of Fornax, that would mean a $r_\mathrm{c}$ of 1304\,pc. To check whether such core size would allow the GCs to survive outside of their $d_\mathrm{p}$, we rerun our orbital evolution model on a finer grid of $r_\mathrm{s}$, $r_\mathrm{c}$ in between 1000\,pc and 2000\,pc, in steps of 100\,pc, with the inclusion of a stellar disk. We find that the minimum required ($r_\mathrm{s}$, $r_\mathrm{c}$) is (1600, 1500)\,pc in order for all GC1 to GC4 to survive outside of their $d_\mathrm{p}$. Such a core size is larger than expected from the coring of the dark matter halo due to baryonic feedback alone in the CDM scenario, given the feedback recipe in \cite{read16}. Observationally, \cite{ber18} have shown that given the star-formation history of Fornax derived by \cite{deB12}, to produce such a large DM core from stellar feedback alone would imply that $\gtrsim30\%$ of that energy is used in the coring of the DM halo, which is $\gtrsim$ two times the maximum fraction of energy from stellar feedback that can be coupled to the retained gas.\footnote{Note that when using the SFH obtained by \cite{delp13}, \cite{ber18} derived a lower required energy fraction of $\sim10\%$ for the creation of a DM core of size $\sim$1.5\,kpc. Although this SFH comes from a deeper photometric data obtained using VLT/FORS (as compared to the CTIO/Mosaic II data used to derive the SFH in \cite{deB12}), its central and less extensive spatial coverage might lead to an overestimation of the overall SFR and hence the total feedback energy.} However given that the merger required for Fornax may also cause some expansion of the DM profile such DM core size might still be possible in the CDM scenario, and should be tested with simulations.

The halo profile constraints may have implications for non-standard DM particle theories as well. With respect to the ultra-light Bose-Einstein condensate dark matter ($\psi$DM), our result can provide constraints on the dark matter particle mass. With cosmological simulations, \cite{sch140} found that the core size of a $\psi$DM halo ($r_\mathrm{c, \psi DM}$) should obey a scaling with the total halo mass $M_\mathrm{vir}$:
\begin{equation}
r_\mathrm{c, \psi DM} = 1.6\,\mathrm{kpc}\Big(\frac{M_\mathrm{vir}}{10^9M_\odot}\Big)^{-1/3}m_{22}^{-1},
\label{eq_rcm22}
\end{equation}
where $m_{22}$ is related to the dark matter particle mass $m_\mathrm{\psi DM}$ as: 
\begin{equation}
m_{22} \equiv\frac{m_\mathrm{\psi DM}}{10^{-22}\,\mathrm{eV}/c^2}.
\end{equation}
From the fitting to the observed $\sigma_\star(R)$, the derived $\rho_\mathrm{c}$ for a dark matter halo of ($r_\mathrm{s}$, $r_\mathrm{c}$) = (1700, 1500)\,pc is 0.03\,$M_\odot$\,pc$^{-3}$. The $M_\mathrm{vir}$ of such a profile is $2.93\times10^9\,M_\odot$\footnote{While the $M_{200}$ of our dark matter halo is $3.24\times10^9\,M_\odot$, here we calculate the virial mass as $M_\mathrm{vir} = (4\pi/3 r_\mathrm{vir}^3)\Delta_\mathrm{c}\rho_\mathrm{c}$ with $\rho_\mathrm{c}$ being the critical density and $\Delta_\mathrm{c} = 350$ following \cite{sch140}}. We have also fitted the derived cNFW density profile with that characterised for $\psi$DM by \cite{sch14} and obtained $r_\mathrm{c, \psi DM}\sim1006$\,pc\footnote{The $\psi$DM density profile is characterised by an inner soliton that transit abruptly to an outer NFW halo. When fitting our derived DM density profile with that of $\psi$DM, we have fixed the transition radius to be 3\,$r_\mathrm{c, \psi DM}$, a cosmic average found by \cite{sch14}. Our derived $r_\mathrm{c, \psi DM}$ is comparable with their derived value of $r_\mathrm{c, \psi DM}=920^{+150}_{-110}$\,pc, found by using the velocity dispersion from three different stellar population in Fornax.}. Using Eq.\ref{eq_rcm22}, we derive a $m_{22}$ of $\sim1.1$, which is within the constraint of $m_{22} = 0.26-2.5$ obtained from large-scale structures \citep[e.g.][]{boz15, sar16}. We note that, however, in the $\psi$DM case, dynamical friction is suppressed by the wave nature of the dark matter particles \citep{hui17} and hence our analysis is not directly applicable. With the suppressed dynamical friction, the required $r_\mathrm{c}$ is likely smaller and hence allows for a larger $m_{22}$. Our work hence still refines the complementary constraints from large-scale structures on $m_{22}$.

In the case of self-interacting dark matter (SIDM), the dark matter halo core size is correlated with the scattering cross-section $\sigma$ as:
\begin{equation}
\frac{\langle\sigma v\rangle}{m_\mathrm{SIDM}}\rho(r_\mathrm{1})t_\mathrm{age}\sim1,
\end{equation}
where $v$ and $m_\mathrm{SIDM}$ is the velocity between the DM particles and the mass of the DM particles, $t_\mathrm{age}$ is the age of the halo, and $r_1$ is the characteristic radius beyond which, the DM particles are scattered less than once per particle on average over $t_\mathrm{age}$ \citep{kap16}. The dark matter halo can be described by an NFW profile beyond $r_1$ and hence this characteristic radius would correspond to the core radius $r_\mathrm{c}$ in the cNFW profiles that we adopted.  $\rho(r_\mathrm{c})$ of the profile with the minimum required ($r_\mathrm{s}$, $r_\mathrm{c}$) of (1600, 1500)\,pc is 0.0095\,$M_\odot\,$pc$^{-3}$, corresponding to a $\frac{\langle\sigma v\rangle}{m_\mathrm{SIDM}}$ of $\sim36\,(\mathrm{cm^2/g}\times\mathrm{km/s})$. Our derived value for Fornax is comparable to other dwarfs or low-surface brightness galaxies in \cite{kap16}.

\subsection{Caveats}

In attempting to incorporate several evolutionary aspects of Fornax in one model, there will necessarily be caveats and simplifications. We outline these here, and hope this work motivates future studies to produce idealised numerical simulations which can test this scenario. When estimating the $d_\mathrm{form}$ of the GCs, we assume a well ordered, exponential disk, while the current structure is much more of a thick oblate blob of stars. Although such a structure could have resulted from the past merger event, in the case where the structure of the stellar component was already puffy when the GCs are formed, we would have overestimated the maximum galactocentric distance at which the GCs can be formed. This is because given the same scale radius and mass, a thicker disc would render a lower density at each specific location. A smaller $d_\mathrm{form}$ would only increase the required ($r_\mathrm{s}, r_\mathrm{c}$) in order for the GCs to survive outside of its present-day $d_\mathrm{p}$ and hence our derived dark matter parameters would still serve as a lower-bound as intended.

Although the underlying dark matter profile is expected to vary due to cosmological halo growth, within the timescales (after the first Gyr since the beginning of the universe) and radial range ($d_\mathrm{form} < 2000$\,pc) relevant for the orbital decay of the GCs, the change of dark matter profile under cosmological halo growth has a negligible impact for our orbit calculations when we tested orbit integration in a growing potential.

While we have assumed a dry merger scenario, the Fornax SFH tells us that stars were still being formed beyond our assumed merger time of $t\sim10$\,Gyr \citep{delp13}. While the SFR clearly drops at $t\sim10$\,Gyr, around half of the stellar mass of Fornax were still formed afterwards, suggesting that gas must be acquired through some other means (if not through mergers) to form those late-forming stars. A possible source could be infall of enriched gas previously expelled through stellar winds or supernova feedbacks. On the other hand, the merger could also have happened later or multiple mergers could have happened. Our model can be applied to arbitrary merging times and/or number of mergers to explore the possible merger histories, but such exploration is beyond the scope of the current paper, as our model on GC orbital evolution alone would not provide power to distinguish between effects of different merger histories and merger mass ratios. The merger number and merger time therefore has to enter as an assumption or from other lines of evidences. We would like to however point out that our derived merger mass ratio is supported by chemical evidence (Section \ref{subsect_mgcfs}). While this certainly does not mean that our proposed scenario is the only possible story, it is a chemodynamically consistent one. Alternatively, the merger could of course also have been semi-wet or wet. In these cases, our modelled dry merger scenario would still act as a lower limit on both the DM core size and the merger mass ratio, as the dry merger scenario provides an upped limit to the possible orbit expansion for a particular merger mass ratio.

Baryonic feedback can additionally cause the coring of the dark matter profile, and lead to the expansion of the GCs' orbit in additional ways. Just like dark matter particles, the GCs gain energy indirectly from stellar feedback ejecting gas in the inner regions of the galaxy and rapidly altering the potential. The repetitive deposition of such energy and subsequent ejection of gas leads to an irreversible non-adiabatic heating of the orbits of the particle in the potential \citep{pon12}. While \cite{pon12} provide analytic expressions for how the overall spatial scale of a system of (e.g. dark matter) particles would be altered given an amount of energy, the effect on an individual particle (e.g. a GC) by such deposition of energy is not well understood and hence not included in our model. Secondly, the GCs would move outwards due to the gradual (rather than instantaneous) shallowing of the gravitational potential. The resultant position of GCs in a coring profile would still lie between the final position under an NFW and a cored profile of the same $r_\mathrm{s}$ and $r_\mathrm{c}$, with that from the cored profile giving an upper bound. The exact position would depend on the timescale for core creation. Since we do not possess information on the timescale at which the dark matter halo change from a cuspy to a cored profile, we only consider the completely cored ($n=1$) cases to obtain an upper limit of the final GCs positions for each set of dark matter parameters ($r_\mathrm{s}, r_\mathrm{c}$).

Lastly, we have also considered a spherical system where both the geometry of the gravitational potential as well as the velocity anisotropy is isotropic. How axisymmetric or triaxial potentials with anisotropic velocity dispersions would affect our result is beyond the scope of this paper. 

\section{Conclusions}\label{sect_con}
We present an analysis on how the present day location of the five globular clusters in the dwarf spheroidal galaxy Fornax provides constraints on its dark matter halo profile. In particular, we incorporate a careful consideration on the formation location of the GCs based on pressure equilibrium arguments, and allow for orbital expansion due to a past merger. We also consider the effect of dynamical buoyancy by including the effect of fast-moving background particles in our dynamical friction treatment, and adopt a velocity distribution function computed self-consistently from each gravitational potential using the Eddington equation (instead of the commonly adopted assumption of a Maxwellian distribution). With these ingredients we study the orbital decay of Fornax's five GCs in a self-consistent framework with their co-evolution of the dynamics and chemistry of the host galaxy. Our main findings from this joint analysis are as follows:

\begin{enumerate}
\item  Our joint analysis shows that survival of three of the GCs (1, 2, 3) in Fornax is possible for halo profiles with minimum scale and core radii of 1700 and 1500 pc respectively -  provided that Fornax has had a merger of mass ratio (1:5 $\leq\eta\leq$ 1:2) in its past.  The younger GC4 can survive in any halo profile provided the same merger occurs, we suggest it may have been triggered during the merger ($\sim$10\,Gyrs ago).

\item GC5 can not survive in a halo unless there is a core radii larger than 6\,kpc (3 times the tidal radius).  As stellar feedback based mechanisms for core creation can not produce a change outside the tidal radius, we posit that GC5 could have been brought in with the merging galaxy to the Fornax host.

\item  Consistent with this, we show that GC5 is unique among the five GCs in that it lies off the Fornax field star age-metallicity relation, with a lower metallicity at fixed age, suggestive of being born in a galaxy with 1/3 the mass of Fornax.

\item This is also supported by empirical chemical evolution arguments. The MDF of Fornax's fields stars are shown to be consistently reproduced by a weighted super-position of pairs of Local Group dwarfs with the necessary mass ratio.  

\item This merger origin for the evolution and survival of Fornax and its GCs reconciles the large number of GCs within Fornax, and alleviates the problem of Fornax having an extremely high mass in GC stars relative to metal poor field stars, as well as its high specific globular cluster frequency of $S_N=29$ \citep{vdb98}.

\item We have compared the required dark matter core size with several dark matter models and find that a dark matter core of 1600\,pc is larger than that expected from baryonic feedback alone in the CDM paradigm. Even though we did not incorporate the wave nature of $\psi$DM in our dynamical friction model, our derived particle mass of $m_{22}\sim$0.7 is still marginally consistent with the lower limit from large-scale structure constraints. Lastly, we find a scattering cross-section of $\frac{\langle\sigma v\rangle}{m_\mathrm{SIDM}}$ of $\sim55\,(\mathrm{cm^2/g}\times\mathrm{km/s})$ for SIDM, consistent with values obtained for other dwarf and low-surface brightness galaxies in the literature.

\end{enumerate}
A putative merger in the evolution of Fornax may support many of the structural, dynamical and stellar population peculiarities it and its GCs show.  A better understanding of the early environment and infall of Fornax within the Local Group will help further understanding of the frequency and impact of low-mass satellite interactions \citep[e.g.][]{star16}. Our suggested scenario, whereby Fornax and its GC populations were assembled by merging dwarfs (with one GC coming in through the merger, one formed during the merger and three pre-exisiting in the proto-Fornax) can be tested with high resolution idealised simulations, and may provide constraints on how common this mechanism is for dwarfs in a cosmological framework.

\section{Acknowledgments}
The authors would like to thank the anonymous referee for a helpful report which improved this manuscript.We would like to thank Else Starkenburg, Chris Brook and Arianna di Cintio for useful discussions which helped improve this manuscript.  RL was supported by funding from the Natural Sciences and Engineering Research Council of Canada PDF award, and this work was supported by Sonderforschungsbereich SFB 881 "The Milky Way System" (subproject A7 and A8) of the Deutsche Forschungsgemeinschaft (DFG), and DAAD PPP project number 57316058 "Finding and exploiting accreted star clusters in the Milky Way". GL and GvdV acknowledge support from the German Academic Exchange Service (DAAD) under PPP project ID 57319730. 
GvdV acknowledges funding from the European Research Council (ERC) under the European Union's Horizon 2020 research and innovation programme under grant agreement No 724857 (Consolidator Grant ArcheoDyn). G.B. gratefully acknowledges financial support from the Spanish Ministry of Economy and Competitiveness (MINECO) under the Ramon y Cajal Programme (RYC-2012-11537) and the grant AYA2017-89076-P.

\bibliographystyle{mnras}
\bibliography{examplerefs}  

\begin{thebibliography}{}
\makeatletter
\relax
\def\mn@urlcharsother{\let\do\@makeother \do\$\do\&\do\#\do\^\do\_\do\%\do\~}
\def\mn@doi{\begingroup\mn@urlcharsother \@ifnextchar [ {\mn@doi@}
  {\mn@doi@[]}}
\def\mn@doi@[#1]#2{\def\@tempa{#1}\ifx\@tempa\@empty \href
  {http://dx.doi.org/#2} {doi:#2}\else \href {http://dx.doi.org/#2} {#1}\fi
  \endgroup}
\def\mn@eprint#1#2{\mn@eprint@#1:#2::\@nil}
\def\mn@eprint@arXiv#1{\href {http://arxiv.org/abs/#1} {{\tt arXiv:#1}}}
\def\mn@eprint@dblp#1{\href {http://dblp.uni-trier.de/rec/bibtex/#1.xml}
  {dblp:#1}}
\def\mn@eprint@#1:#2:#3:#4\@nil{\def\@tempa {#1}\def\@tempb {#2}\def\@tempc
  {#3}\ifx \@tempc \@empty \let \@tempc \@tempb \let \@tempb \@tempa \fi \ifx
  \@tempb \@empty \def\@tempb {arXiv}\fi \@ifundefined
  {mn@eprint@\@tempb}{\@tempb:\@tempc}{\expandafter \expandafter \csname
  mn@eprint@\@tempb\endcsname \expandafter{\@tempc}}}

\bibitem[\protect\citeauthoryear{{Adams} et~al.,}{{Adams}
  et~al.}{2014}]{adams14}
{Adams} J.~J.,  et~al., 2014, \mn@doi [\apj] {10.1088/0004-637X/789/1/63},
  \href {http://adsabs.harvard.edu/abs/2014ApJ...789...63A} {789, 63}

\bibitem[\protect\citeauthoryear{{Amorisco}}{{Amorisco}}{2017}]{amo17}
{Amorisco} N.~C.,  2017, \mn@doi [\apj] {10.3847/1538-4357/aa745f}, \href
  {http://adsabs.harvard.edu/abs/2017ApJ...844...64A} {844, 64}

\bibitem[\protect\citeauthoryear{{Amorisco} \& {Evans}}{{Amorisco} \&
  {Evans}}{2011}]{amo11}
{Amorisco} N.~C.,  {Evans} N.~W.,  2011, \mn@doi [\mnras]
  {10.1111/j.1365-2966.2010.17715.x}, \href
  {http://adsabs.harvard.edu/abs/2011MNRAS.411.2118A} {411, 2118}

\bibitem[\protect\citeauthoryear{{Amorisco} \& {Evans}}{{Amorisco} \&
  {Evans}}{2012}]{amo12}
{Amorisco} N.~C.,  {Evans} N.~W.,  2012, \mn@doi [\apjl]
  {10.1088/2041-8205/756/1/L2}, \href
  {http://adsabs.harvard.edu/abs/2012ApJ...756L...2A} {756, L2}

\bibitem[\protect\citeauthoryear{{Amorisco}, {Agnello}  \& {Evans}}{{Amorisco}
  et~al.}{2013}]{amo13}
{Amorisco} N.~C.,  {Agnello} A.,   {Evans} N.~W.,  2013, \mn@doi [\mnras]
  {10.1093/mnrasl/sls031}, \href
  {http://adsabs.harvard.edu/abs/2013MNRAS.429L..89A} {429, L89}

\bibitem[\protect\citeauthoryear{{Amorisco}, {Evans}  \& {van de
  Ven}}{{Amorisco} et~al.}{2014}]{amo14}
{Amorisco} N.~C.,  {Evans} N.~W.,   {van de Ven} G.,  2014, \mn@doi [\nat]
  {10.1038/nature12995}, \href
  {http://adsabs.harvard.edu/abs/2014Natur.507..335A} {507, 335}

\bibitem[\protect\citeauthoryear{{Angus} \& {Diaferio}}{{Angus} \&
  {Diaferio}}{2009}]{an09}
{Angus} G.~W.,  {Diaferio} A.,  2009, \mn@doi [\mnras]
  {10.1111/j.1365-2966.2009.14745.x}, \href
  {http://adsabs.harvard.edu/abs/2009MNRAS.396..887A} {396, 887}

\bibitem[\protect\citeauthoryear{{Arca-Sedda} \&
  {Capuzzo-Dolcetta}}{{Arca-Sedda} \& {Capuzzo-Dolcetta}}{2016}]{arca16}
{Arca-Sedda} M.,  {Capuzzo-Dolcetta} R.,  2016, \mn@doi [\mnras]
  {10.1093/mnras/stw1647}, \href
  {http://adsabs.harvard.edu/abs/2016MNRAS.461.4335A} {461, 4335}

\bibitem[\protect\citeauthoryear{{Bastian}}{{Bastian}}{2017}]{bas17}
{Bastian} N.,  2017, in {Charbonnel} C.,  {Nota} A.,  eds,  IAU Symposium Vol.
  316, Formation, Evolution, and Survival of Massive Star Clusters. pp
  302--309, \mn@doi{10.1017/S1743921315009692}

\bibitem[\protect\citeauthoryear{{Battaglia} et~al.,}{{Battaglia}
  et~al.}{2006}]{bat06}
{Battaglia} G.,  et~al., 2006, \mn@doi [\aap] {10.1051/0004-6361:20065720},
  \href {http://adsabs.harvard.edu/abs/2006A%26A...459..423B} {459, 423}

\bibitem[\protect\citeauthoryear{{Battaglia}, {Helmi}, {Tolstoy}, {Irwin},
  {Hill}  \& {Jablonka}}{{Battaglia} et~al.}{2008}]{bat08}
{Battaglia} G.,  {Helmi} A.,  {Tolstoy} E.,  {Irwin} M.,  {Hill} V.,
  {Jablonka} P.,  2008, \mn@doi [\apjl] {10.1086/590179}, \href
  {http://adsabs.harvard.edu/abs/2008ApJ...681L..13B} {681, L13}

\bibitem[\protect\citeauthoryear{{Battaglia}, {Tolstoy}, {Helmi}, {Irwin},
  {Parisi}, {Hill}  \& {Jablonka}}{{Battaglia} et~al.}{2011}]{bat11}
{Battaglia} G.,  {Tolstoy} E.,  {Helmi} A.,  {Irwin} M.,  {Parisi} P.,  {Hill}
  V.,   {Jablonka} P.,  2011, \mn@doi [\mnras]
  {10.1111/j.1365-2966.2010.17745.x}, \href
  {http://adsabs.harvard.edu/abs/2011MNRAS.411.1013B} {411, 1013}

\bibitem[\protect\citeauthoryear{{Bechtol} et~al.,}{{Bechtol}
  et~al.}{2015}]{bech15}
{Bechtol} K.,  et~al., 2015, \mn@doi [\apj] {10.1088/0004-637X/807/1/50}, \href
  {http://adsabs.harvard.edu/abs/2015ApJ...807...50B} {807, 50}

\bibitem[\protect\citeauthoryear{{Ben{\'{\i}}tez-Llambay}, {Navarro}, {Abadi},
  {Gottl{\"o}ber}, {Yepes}, {Hoffman}  \& {Steinmetz}}{{Ben{\'{\i}}tez-Llambay}
  et~al.}{2016}]{ben16}
{Ben{\'{\i}}tez-Llambay} A.,  {Navarro} J.~F.,  {Abadi} M.~G.,  {Gottl{\"o}ber}
  S.,  {Yepes} G.,  {Hoffman} Y.,   {Steinmetz} M.,  2016, \mn@doi [\mnras]
  {10.1093/mnras/stv2722}, \href
  {http://adsabs.harvard.edu/abs/2016MNRAS.456.1185B} {456, 1185}

\bibitem[\protect\citeauthoryear{{Bermejo-Climent} et~al.,}{{Bermejo-Climent}
  et~al.}{2018}]{ber18}
{Bermejo-Climent} J.~R.,  et~al., 2018, \mn@doi [\mnras]
  {10.1093/mnras/sty1651}, \href
  {http://adsabs.harvard.edu/abs/2018MNRAS.479.1514B} {479, 1514}

\bibitem[\protect\citeauthoryear{{Binney} \& {Tremaine}}{{Binney} \&
  {Tremaine}}{1987}]{bin87}
{Binney} J.,  {Tremaine} S.,  1987, {Galactic dynamics}

\bibitem[\protect\citeauthoryear{{Bozek}, {Marsh}, {Silk}  \& {Wyse}}{{Bozek}
  et~al.}{2015}]{boz15}
{Bozek} B.,  {Marsh} D.~J.~E.,  {Silk} J.,   {Wyse} R.~F.~G.,  2015, \mn@doi
  [\mnras] {10.1093/mnras/stv624}, \href
  {http://adsabs.harvard.edu/abs/2015MNRAS.450..209B} {450, 209}

\bibitem[\protect\citeauthoryear{{Brook}}{{Brook}}{2015}]{bro15}
{Brook} C.~B.,  2015, \mn@doi [\mnras] {10.1093/mnras/stv2101}, \href
  {http://adsabs.harvard.edu/abs/2015MNRAS.454.1719B} {454, 1719}

\bibitem[\protect\citeauthoryear{{Burkert}}{{Burkert}}{2015}]{bur15}
{Burkert} A.,  2015, \mn@doi [\apj] {10.1088/0004-637X/808/2/158}, \href
  {http://adsabs.harvard.edu/abs/2015ApJ...808..158B} {808, 158}

\bibitem[\protect\citeauthoryear{{Chandrasekhar}}{{Chandrasekhar}}{1943}]{chan43}
{Chandrasekhar} S.,  1943, \mn@doi [\apj] {10.1086/144517}, \href
  {http://adsabs.harvard.edu/abs/1943ApJ....97..255C} {97, 255}

\bibitem[\protect\citeauthoryear{{Cicu{\'e}ndez} \&
  {Battaglia}}{{Cicu{\'e}ndez} \& {Battaglia}}{2018}]{cic18}
{Cicu{\'e}ndez} L.,  {Battaglia} G.,  2018, \mn@doi [\mnras]
  {10.1093/mnras/sty1748}, \href
  {http://adsabs.harvard.edu/abs/2018MNRAS.480..251C} {480, 251}

\bibitem[\protect\citeauthoryear{{Cole}, {Dehnen}  \& {Wilkinson}}{{Cole}
  et~al.}{2011}]{cole11}
{Cole} D.~R.,  {Dehnen} W.,   {Wilkinson} M.~I.,  2011, \mn@doi [\mnras]
  {10.1111/j.1365-2966.2011.19110.x}, \href
  {http://adsabs.harvard.edu/abs/2011MNRAS.416.1118C} {416, 1118}

\bibitem[\protect\citeauthoryear{{Cole}, {Dehnen}, {Read}  \&
  {Wilkinson}}{{Cole} et~al.}{2012}]{cole12}
{Cole} D.~R.,  {Dehnen} W.,  {Read} J.~I.,   {Wilkinson} M.~I.,  2012, \mn@doi
  [\mnras] {10.1111/j.1365-2966.2012.21885.x}, \href
  {http://adsabs.harvard.edu/abs/2012MNRAS.426..601C} {426, 601}

\bibitem[\protect\citeauthoryear{{Coleman} \& {de Jong}}{{Coleman} \& {de
  Jong}}{2008}]{col08}
{Coleman} M.~G.,  {de Jong} J.~T.~A.,  2008, \mn@doi [\apj] {10.1086/589992},
  \href {http://adsabs.harvard.edu/abs/2008ApJ...685..933C} {685, 933}

\bibitem[\protect\citeauthoryear{{Coleman}, {Da Costa}, {Bland-Hawthorn}  \&
  {Freeman}}{{Coleman} et~al.}{2005}]{col05}
{Coleman} M.~G.,  {Da Costa} G.~S.,  {Bland-Hawthorn} J.,   {Freeman} K.~C.,
  2005, \mn@doi [\aj] {10.1086/427966}, \href
  {http://adsabs.harvard.edu/abs/2005AJ....129.1443C} {129, 1443}

\bibitem[\protect\citeauthoryear{{Contenta} et~al.,}{{Contenta}
  et~al.}{2018}]{con18}
{Contenta} F.,  et~al., 2018, \mn@doi [\mnras] {10.1093/mnras/sty424}, \href
  {http://adsabs.harvard.edu/abs/2018MNRAS.476.3124C} {476, 3124}

\bibitem[\protect\citeauthoryear{{Cowsik}, {Wagoner}, {Berti}  \&
  {Sircar}}{{Cowsik} et~al.}{2009}]{cos09}
{Cowsik} R.,  {Wagoner} K.,  {Berti} E.,   {Sircar} A.,  2009, \mn@doi [\apj]
  {10.1088/0004-637X/699/2/1389}, \href
  {http://adsabs.harvard.edu/abs/2009ApJ...699.1389C} {699, 1389}

\bibitem[\protect\citeauthoryear{{Dutton} \& {Macci{\`o}}}{{Dutton} \&
  {Macci{\`o}}}{2014}]{dut14}
{Dutton} A.~A.,  {Macci{\`o}} A.~V.,  2014, \mn@doi [\mnras]
  {10.1093/mnras/stu742}, \href
  {http://adsabs.harvard.edu/abs/2014MNRAS.441.3359D} {441, 3359}

\bibitem[\protect\citeauthoryear{{Dutton} \& {van den Bosch}}{{Dutton} \& {van
  den Bosch}}{2009}]{dut09}
{Dutton} A.~A.,  {van den Bosch} F.~C.,  2009, \mn@doi [\mnras]
  {10.1111/j.1365-2966.2009.14742.x}, \href
  {http://adsabs.harvard.edu/abs/2009MNRAS.396..141D} {396, 141}

\bibitem[\protect\citeauthoryear{{Elmegreen} \& {Efremov}}{{Elmegreen} \&
  {Efremov}}{1997}]{elm97}
{Elmegreen} B.~G.,  {Efremov} Y.~N.,  1997, \mn@doi [\apj] {10.1086/303966},
  \href {http://adsabs.harvard.edu/abs/1997ApJ...480..235E} {480, 235}

\bibitem[\protect\citeauthoryear{{Evans} \& {An}}{{Evans} \& {An}}{2006}]{ev06}
{Evans} N.~W.,  {An} J.~H.,  2006, \mn@doi [\prd] {10.1103/PhysRevD.73.023524},
  \href {http://adsabs.harvard.edu/abs/2006PhRvD..73b3524E} {73, 023524}

\bibitem[\protect\citeauthoryear{{Faber} \& {Jackson}}{{Faber} \&
  {Jackson}}{1976}]{fab76}
{Faber} S.~M.,  {Jackson} R.~E.,  1976, \mn@doi [\apj] {10.1086/154215}, \href
  {http://adsabs.harvard.edu/abs/1976ApJ...204..668F} {204, 668}

\bibitem[\protect\citeauthoryear{{Gaia Collaboration} et~al.,}{{Gaia
  Collaboration} et~al.}{2018}]{hel18}
{Gaia Collaboration} et~al., 2018, preprint, \href
  {http://adsabs.harvard.edu/abs/2018arXiv180409381G} {} (\mn@eprint {arXiv}
  {1804.09381})

\bibitem[\protect\citeauthoryear{{Goerdt}, {Moore}, {Read}, {Stadel}  \&
  {Zemp}}{{Goerdt} et~al.}{2006}]{goe06}
{Goerdt} T.,  {Moore} B.,  {Read} J.~I.,  {Stadel} J.,   {Zemp} M.,  2006,
  \mn@doi [\mnras] {10.1111/j.1365-2966.2006.10182.x}, \href
  {http://adsabs.harvard.edu/abs/2006MNRAS.368.1073G} {368, 1073}

\bibitem[\protect\citeauthoryear{{Hansen}, {Egli}, {Hollenstein}  \&
  {Salzmann}}{{Hansen} et~al.}{2005}]{han05}
{Hansen} S.~H.,  {Egli} D.,  {Hollenstein} L.,   {Salzmann} C.,  2005, \mn@doi
  [\na] {10.1016/j.newast.2005.01.005}, \href
  {http://adsabs.harvard.edu/abs/2005NewA...10..379H} {10, 379}

\bibitem[\protect\citeauthoryear{{Hernandez} \& {Gilmore}}{{Hernandez} \&
  {Gilmore}}{1998}]{her98}
{Hernandez} X.,  {Gilmore} G.,  1998, \mn@doi [\mnras]
  {10.1046/j.1365-8711.1998.01511.x}, \href
  {http://adsabs.harvard.edu/abs/1998MNRAS.297..517H} {297, 517}

\bibitem[\protect\citeauthoryear{{Hui}, {Ostriker}, {Tremaine}  \&
  {Witten}}{{Hui} et~al.}{2017}]{hui17}
{Hui} L.,  {Ostriker} J.~P.,  {Tremaine} S.,   {Witten} E.,  2017, \mn@doi
  [\prd] {10.1103/PhysRevD.95.043541}, \href
  {http://adsabs.harvard.edu/abs/2017PhRvD..95d3541H} {95, 043541}

\bibitem[\protect\citeauthoryear{{Inoue}}{{Inoue}}{2009}]{in09}
{Inoue} S.,  2009, \mn@doi [\mnras] {10.1111/j.1365-2966.2009.15066.x}, \href
  {http://adsabs.harvard.edu/abs/2009MNRAS.397..709I} {397, 709}

\bibitem[\protect\citeauthoryear{{Inoue}}{{Inoue}}{2011}]{in11}
{Inoue} S.,  2011, \mn@doi [\mnras] {10.1111/j.1365-2966.2011.19122.x}, \href
  {http://adsabs.harvard.edu/abs/2011MNRAS.416.1181I} {416, 1181}

\bibitem[\protect\citeauthoryear{{Jackson}, {Skillman}, {Gehrz}, {Polomski}  \&
  {Woodward}}{{Jackson} et~al.}{2007}]{jack07}
{Jackson} D.~C.,  {Skillman} E.~D.,  {Gehrz} R.~D.,  {Polomski} E.,
  {Woodward} C.~E.,  2007, \mn@doi [\apj] {10.1086/510354}, \href
  {http://adsabs.harvard.edu/abs/2007ApJ...656..818J} {656, 818}

\bibitem[\protect\citeauthoryear{{Kacharov} et~al.,}{{Kacharov}
  et~al.}{2017}]{kach17}
{Kacharov} N.,  et~al., 2017, \mn@doi [\mnras] {10.1093/mnras/stw3188}, \href
  {http://adsabs.harvard.edu/abs/2017MNRAS.466.2006K} {466, 2006}

\bibitem[\protect\citeauthoryear{{Kaplinghat}, {Tulin}  \& {Yu}}{{Kaplinghat}
  et~al.}{2016}]{kap16}
{Kaplinghat} M.,  {Tulin} S.,   {Yu} H.-B.,  2016, \mn@doi [Physical Review
  Letters] {10.1103/PhysRevLett.116.041302}, \href
  {http://adsabs.harvard.edu/abs/2016PhRvL.116d1302K} {116, 041302}

\bibitem[\protect\citeauthoryear{{Kirby}, {Cohen}, {Guhathakurta}, {Cheng},
  {Bullock}  \& {Gallazzi}}{{Kirby} et~al.}{2013}]{kir13}
{Kirby} E.~N.,  {Cohen} J.~G.,  {Guhathakurta} P.,  {Cheng} L.,  {Bullock}
  J.~S.,   {Gallazzi} A.,  2013, \mn@doi [\apj] {10.1088/0004-637X/779/2/102},
  \href {http://adsabs.harvard.edu/abs/2013ApJ...779..102K} {779, 102}

\bibitem[\protect\citeauthoryear{{Koposov}, {Belokurov}, {Torrealba}  \&
  {Evans}}{{Koposov} et~al.}{2015}]{kop15}
{Koposov} S.~E.,  {Belokurov} V.,  {Torrealba} G.,   {Evans} N.~W.,  2015,
  \mn@doi [\apj] {10.1088/0004-637X/805/2/130}, \href
  {http://adsabs.harvard.edu/abs/2015ApJ...805..130K} {805, 130}

\bibitem[\protect\citeauthoryear{{Kourkchi} et~al.,}{{Kourkchi}
  et~al.}{2012}]{kou12}
{Kourkchi} E.,  et~al., 2012, \mn@doi [\mnras]
  {10.1111/j.1365-2966.2011.19899.x}, \href
  {http://adsabs.harvard.edu/abs/2012MNRAS.420.2819K} {420, 2819}

\bibitem[\protect\citeauthoryear{{Kruijssen}}{{Kruijssen}}{2015}]{kru15}
{Kruijssen} J.~M.~D.,  2015, \mn@doi [\mnras] {10.1093/mnras/stv2026}, \href
  {http://adsabs.harvard.edu/abs/2015MNRAS.454.1658K} {454, 1658}

\bibitem[\protect\citeauthoryear{{Krumholz} \& {McKee}}{{Krumholz} \&
  {McKee}}{2005}]{kru05}
{Krumholz} M.~R.,  {McKee} C.~F.,  2005, \mn@doi [\apj] {10.1086/431734}, \href
  {http://adsabs.harvard.edu/abs/2005ApJ...630..250K} {630, 250}

\bibitem[\protect\citeauthoryear{{Kuhlen}, {Weiner}, {Diemand}, {Madau},
  {Moore}, {Potter}, {Stadel}  \& {Zemp}}{{Kuhlen} et~al.}{2010}]{kuh10}
{Kuhlen} M.,  {Weiner} N.,  {Diemand} J.,  {Madau} P.,  {Moore} B.,  {Potter}
  D.,  {Stadel} J.,   {Zemp} M.,  2010, \mn@doi [\jcap]
  {10.1088/1475-7516/2010/02/030}, \href
  {http://adsabs.harvard.edu/abs/2010JCAP...02..030K} {2, 030}

\bibitem[\protect\citeauthoryear{{Larsen}, {Strader}  \& {Brodie}}{{Larsen}
  et~al.}{2012}]{lar12}
{Larsen} S.~S.,  {Strader} J.,   {Brodie} J.~P.,  2012, \mn@doi [\aap]
  {10.1051/0004-6361/201219897}, \href
  {http://adsabs.harvard.edu/abs/2012A%26A...544L..14L} {544, L14}

\bibitem[\protect\citeauthoryear{{Leaman} et~al.,}{{Leaman}
  et~al.}{2013}]{lea13}
{Leaman} R.,  et~al., 2013, \mn@doi [\apj] {10.1088/0004-637X/767/2/131}, \href
  {http://adsabs.harvard.edu/abs/2013ApJ...767..131L} {767, 131}

\bibitem[\protect\citeauthoryear{{Leaman} et~al.,}{{Leaman}
  et~al.}{2017}]{lea17}
{Leaman} R.,  et~al., 2017, \mn@doi [\mnras] {10.1093/mnras/stx2014}, \href
  {http://adsabs.harvard.edu/abs/2017MNRAS.472.1879L} {472, 1879}

\bibitem[\protect\citeauthoryear{{Lima Neto}, {Gerbal}  \& {M{\'a}rquez}}{{Lima
  Neto} et~al.}{1999}]{lim99}
{Lima Neto} G.~B.,  {Gerbal} D.,   {M{\'a}rquez} I.,  1999, \mn@doi [\mnras]
  {10.1046/j.1365-8711.1999.02849.x}, \href
  {http://adsabs.harvard.edu/abs/1999MNRAS.309..481L} {309, 481}

\bibitem[\protect\citeauthoryear{{Lovell}, {Frenk}, {Eke}, {Jenkins}, {Gao}  \&
  {Theuns}}{{Lovell} et~al.}{2014}]{lov14}
{Lovell} M.~R.,  {Frenk} C.~S.,  {Eke} V.~R.,  {Jenkins} A.,  {Gao} L.,
  {Theuns} T.,  2014, \mn@doi [\mnras] {10.1093/mnras/stt2431}, \href
  {http://adsabs.harvard.edu/abs/2014MNRAS.439..300L} {439, 300}

\bibitem[\protect\citeauthoryear{{Lux}, {Read}  \& {Lake}}{{Lux}
  et~al.}{2010}]{lux10}
{Lux} H.,  {Read} J.~I.,   {Lake} G.,  2010, \mn@doi [\mnras]
  {10.1111/j.1365-2966.2010.16877.x}, \href
  {http://adsabs.harvard.edu/abs/2010MNRAS.406.2312L} {406, 2312}

\bibitem[\protect\citeauthoryear{{Mackey} \& {Gilmore}}{{Mackey} \&
  {Gilmore}}{2003}]{mac03m}
{Mackey} A.~D.,  {Gilmore} G.~F.,  2003, \mn@doi [\mnras]
  {10.1046/j.1365-8711.2003.06275.x}, \href
  {http://adsabs.harvard.edu/abs/2003MNRAS.340..175M} {340, 175}

\bibitem[\protect\citeauthoryear{{Martin} et~al.,}{{Martin}
  et~al.}{2015}]{mar15}
{Martin} N.~F.,  et~al., 2015, \mn@doi [\apjl] {10.1088/2041-8205/804/1/L5},
  \href {http://adsabs.harvard.edu/abs/2015ApJ...804L...5M} {804, L5}

\bibitem[\protect\citeauthoryear{{McConnachie}}{{McConnachie}}{2012}]{mc12}
{McConnachie} A.~W.,  2012, \mn@doi [\aj] {10.1088/0004-6256/144/1/4}, \href
  {http://adsabs.harvard.edu/abs/2012AJ....144....4M} {144, 4}

\bibitem[\protect\citeauthoryear{{Moster}, {Somerville}, {Maulbetsch}, {van den
  Bosch}, {Macci{\`o}}, {Naab}  \& {Oser}}{{Moster} et~al.}{2010}]{mos10}
{Moster} B.~P.,  {Somerville} R.~S.,  {Maulbetsch} C.,  {van den Bosch} F.~C.,
  {Macci{\`o}} A.~V.,  {Naab} T.,   {Oser} L.,  2010, \mn@doi [\apj]
  {10.1088/0004-637X/710/2/903}, \href
  {http://adsabs.harvard.edu/abs/2010ApJ...710..903M} {710, 903}

\bibitem[\protect\citeauthoryear{{Naab}, {Johansson}  \& {Ostriker}}{{Naab}
  et~al.}{2009}]{naab09}
{Naab} T.,  {Johansson} P.~H.,   {Ostriker} J.~P.,  2009, \mn@doi [\apjl]
  {10.1088/0004-637X/699/2/L178}, \href
  {http://adsabs.harvard.edu/abs/2009ApJ...699L.178N} {699, L178}

\bibitem[\protect\citeauthoryear{{Navarro}, {Frenk}  \& {White}}{{Navarro}
  et~al.}{1996}]{nfw96}
{Navarro} J.~F.,  {Frenk} C.~S.,   {White} S.~D.~M.,  1996, \mn@doi [\apj]
  {10.1086/177173}, \href {http://adsabs.harvard.edu/abs/1996ApJ...462..563N}
  {462, 563}

\bibitem[\protect\citeauthoryear{{Oh}, {Lin}  \& {Richer}}{{Oh}
  et~al.}{2000}]{oh00}
{Oh} K.~S.,  {Lin} D.~N.~C.,   {Richer} H.~B.,  2000, \mn@doi [\apj]
  {10.1086/308477}, \href {http://adsabs.harvard.edu/abs/2000ApJ...531..727O}
  {531, 727}

\bibitem[\protect\citeauthoryear{{Oh}, {Brook}, {Governato}, {Brinks}, {Mayer},
  {de Blok}, {Brooks}  \& {Walter}}{{Oh} et~al.}{2011}]{oh11}
{Oh} S.-H.,  {Brook} C.,  {Governato} F.,  {Brinks} E.,  {Mayer} L.,  {de Blok}
  W.~J.~G.,  {Brooks} A.,   {Walter} F.,  2011, \mn@doi [\aj]
  {10.1088/0004-6256/142/1/24}, \href
  {http://adsabs.harvard.edu/abs/2011AJ....142...24O} {142, 24}

\bibitem[\protect\citeauthoryear{{Pe{\~n}arrubia}, {Pontzen}, {Walker}  \&
  {Koposov}}{{Pe{\~n}arrubia} et~al.}{2012}]{pen12}
{Pe{\~n}arrubia} J.,  {Pontzen} A.,  {Walker} M.~G.,   {Koposov} S.~E.,  2012,
  \mn@doi [\apjl] {10.1088/2041-8205/759/2/L42}, \href
  {http://adsabs.harvard.edu/abs/2012ApJ...759L..42P} {759, L42}

\bibitem[\protect\citeauthoryear{{Petts}, {Gualandris}  \& {Read}}{{Petts}
  et~al.}{2015}]{petts15}
{Petts} J.~A.,  {Gualandris} A.,   {Read} J.~I.,  2015, \mn@doi [\mnras]
  {10.1093/mnras/stv2235}, \href
  {http://adsabs.harvard.edu/abs/2015MNRAS.454.3778P} {454, 3778}

\bibitem[\protect\citeauthoryear{{Petts}, {Read}  \& {Gualandris}}{{Petts}
  et~al.}{2016}]{petts16}
{Petts} J.~A.,  {Read} J.~I.,   {Gualandris} A.,  2016, \mn@doi [\mnras]
  {10.1093/mnras/stw2011}, \href
  {http://adsabs.harvard.edu/abs/2016MNRAS.463..858P} {463, 858}

\bibitem[\protect\citeauthoryear{{Pontzen} \& {Governato}}{{Pontzen} \&
  {Governato}}{2012}]{pon12}
{Pontzen} A.,  {Governato} F.,  2012, \mn@doi [\mnras]
  {10.1111/j.1365-2966.2012.20571.x}, \href
  {http://adsabs.harvard.edu/abs/2012MNRAS.421.3464P} {421, 3464}

\bibitem[\protect\citeauthoryear{{Read}, {Goerdt}, {Moore}, {Pontzen}, {Stadel}
   \& {Lake}}{{Read} et~al.}{2006}]{read06}
{Read} J.~I.,  {Goerdt} T.,  {Moore} B.,  {Pontzen} A.~P.,  {Stadel} J.,
  {Lake} G.,  2006, \mn@doi [\mnras] {10.1111/j.1365-2966.2006.11022.x}, \href
  {http://adsabs.harvard.edu/abs/2006MNRAS.373.1451R} {373, 1451}

\bibitem[\protect\citeauthoryear{{Read}, {Agertz}  \& {Collins}}{{Read}
  et~al.}{2016}]{read16}
{Read} J.~I.,  {Agertz} O.,   {Collins} M.~L.~M.,  2016, \mn@doi [\mnras]
  {10.1093/mnras/stw713}, \href
  {http://adsabs.harvard.edu/abs/2016MNRAS.459.2573R} {459, 2573}

\bibitem[\protect\citeauthoryear{{S{\'a}nchez-Salcedo}, {Reyes-Iturbide}  \&
  {Hernandez}}{{S{\'a}nchez-Salcedo} et~al.}{2006}]{san06}
{S{\'a}nchez-Salcedo} F.~J.,  {Reyes-Iturbide} J.,   {Hernandez} X.,  2006,
  \mn@doi [\mnras] {10.1111/j.1365-2966.2006.10602.x}, \href
  {http://adsabs.harvard.edu/abs/2006MNRAS.370.1829S} {370, 1829}

\bibitem[\protect\citeauthoryear{{Sarkar}, {Mondal}, {Das}, {Sethi},
  {Bharadwaj}  \& {Marsh}}{{Sarkar} et~al.}{2016}]{sar16}
{Sarkar} A.,  {Mondal} R.,  {Das} S.,  {Sethi} S.~K.,  {Bharadwaj} S.,
  {Marsh} D.~J.~E.,  2016, \mn@doi [\jcap] {10.1088/1475-7516/2016/04/012},
  \href {http://adsabs.harvard.edu/abs/2016JCAP...04..012S} {4, 012}

\bibitem[\protect\citeauthoryear{{Schive}, {Chiueh}  \& {Broadhurst}}{{Schive}
  et~al.}{2014a}]{sch14}
{Schive} H.-Y.,  {Chiueh} T.,   {Broadhurst} T.,  2014a, \mn@doi [Nature
  Physics] {10.1038/nphys2996}, \href
  {http://adsabs.harvard.edu/abs/2014NatPh..10..496S} {10, 496}

\bibitem[\protect\citeauthoryear{{Schive}, {Liao}, {Woo}, {Wong}, {Chiueh},
  {Broadhurst}  \& {Hwang}}{{Schive} et~al.}{2014b}]{sch140}
{Schive} H.-Y.,  {Liao} M.-H.,  {Woo} T.-P.,  {Wong} S.-K.,  {Chiueh} T.,
  {Broadhurst} T.,   {Hwang} W.-Y.~P.,  2014b, \mn@doi [Physical Review
  Letters] {10.1103/PhysRevLett.113.261302}, \href
  {http://adsabs.harvard.edu/abs/2014PhRvL.113z1302S} {113, 261302}

\bibitem[\protect\citeauthoryear{{Starkenburg} et~al.,}{{Starkenburg}
  et~al.}{2010}]{star10}
{Starkenburg} E.,  et~al., 2010, \mn@doi [\aap] {10.1051/0004-6361/200913759},
  \href {http://adsabs.harvard.edu/abs/2010A%26A...513A..34S} {513, A34}

\bibitem[\protect\citeauthoryear{{Starkenburg}, {Helmi}  \&
  {Sales}}{{Starkenburg} et~al.}{2016}]{star16}
{Starkenburg} T.~K.,  {Helmi} A.,   {Sales} L.~V.,  2016, \mn@doi [\aap]
  {10.1051/0004-6361/201528066}, \href
  {https://ui.adsabs.harvard.edu/abs/2016A&A...595A..56S} {595, A56}

\bibitem[\protect\citeauthoryear{{Stierwalt}, {Besla}, {Patton}, {Johnson},
  {Kallivayalil}, {Putman}, {Privon}  \& {Ross}}{{Stierwalt}
  et~al.}{2015}]{stier15}
{Stierwalt} S.,  {Besla} G.,  {Patton} D.,  {Johnson} K.,  {Kallivayalil} N.,
  {Putman} M.,  {Privon} G.,   {Ross} G.,  2015, \mn@doi [\apj]
  {10.1088/0004-637X/805/1/2}, \href
  {http://adsabs.harvard.edu/abs/2015ApJ...805....2S} {805, 2}

\bibitem[\protect\citeauthoryear{{Tolstoy} et~al.,}{{Tolstoy}
  et~al.}{2004}]{tol04}
{Tolstoy} E.,  et~al., 2004, \mn@doi [\apjl] {10.1086/427388}, \href
  {http://adsabs.harvard.edu/abs/2004ApJ...617L.119T} {617, L119}

\bibitem[\protect\citeauthoryear{{Tremaine}}{{Tremaine}}{1976}]{tre76}
{Tremaine} S.~D.,  1976, \mn@doi [\apj] {10.1086/154085}, \href
  {http://adsabs.harvard.edu/abs/1976ApJ...203..345T} {203, 345}

\bibitem[\protect\citeauthoryear{{Walker} \& {Pe{\~n}arrubia}}{{Walker} \&
  {Pe{\~n}arrubia}}{2011}]{walk11}
{Walker} M.~G.,  {Pe{\~n}arrubia} J.,  2011, \mn@doi [\apj]
  {10.1088/0004-637X/742/1/20}, \href
  {http://adsabs.harvard.edu/abs/2011ApJ...742...20W} {742, 20}

\bibitem[\protect\citeauthoryear{{Walker}, {Mateo}  \& {Olszewski}}{{Walker}
  et~al.}{2009}]{walk09}
{Walker} M.~G.,  {Mateo} M.,   {Olszewski} E.~W.,  2009, \mn@doi [\aj]
  {10.1088/0004-6256/137/2/3100}, \href
  {http://adsabs.harvard.edu/abs/2009AJ....137.3100W} {137, 3100}

\bibitem[\protect\citeauthoryear{{Wang} et~al.,}{{Wang} et~al.}{2019}]{wang19}
{Wang} M.-Y.,  et~al., 2019, arXiv e-prints, \href
  {http://adsabs.harvard.edu/abs/2019arXiv190204589W} {}

\bibitem[\protect\citeauthoryear{{Wetzel}, {Deason}  \&
  {Garrison-Kimmel}}{{Wetzel} et~al.}{2015}]{wet15}
{Wetzel} A.~R.,  {Deason} A.~J.,   {Garrison-Kimmel} S.,  2015, \mn@doi [\apj]
  {10.1088/0004-637X/807/1/49}, \href
  {http://adsabs.harvard.edu/abs/2015ApJ...807...49W} {807, 49}

\bibitem[\protect\citeauthoryear{{Zhu}, {van de Ven}, {Watkins}  \&
  {Posti}}{{Zhu} et~al.}{2016}]{zhu16}
{Zhu} L.,  {van de Ven} G.,  {Watkins} L.~L.,   {Posti} L.,  2016, \mn@doi
  [\mnras] {10.1093/mnras/stw2081}, \href
  {http://adsabs.harvard.edu/abs/2016MNRAS.463.1117Z} {463, 1117}

\bibitem[\protect\citeauthoryear{{de Boer} \& {Fraser}}{{de Boer} \&
  {Fraser}}{2016}]{deB16}
{de Boer} T.~J.~L.,  {Fraser} M.,  2016, \mn@doi [\aap]
  {10.1051/0004-6361/201527580}, \href
  {http://adsabs.harvard.edu/abs/2016A%26A...590A..35D} {590, A35}

\bibitem[\protect\citeauthoryear{{de Boer} et~al.,}{{de Boer}
  et~al.}{2012}]{deB12}
{de Boer} T.~J.~L.,  et~al., 2012, \mn@doi [\aap]
  {10.1051/0004-6361/201219547}, \href
  {http://adsabs.harvard.edu/abs/2012A%26A...544A..73D} {544, A73}

\bibitem[\protect\citeauthoryear{{del Pino}, {Hidalgo}, {Aparicio}, {Gallart},
  {Carrera}, {Monelli}, {Buonanno}  \& {Marconi}}{{del Pino}
  et~al.}{2013}]{delp13}
{del Pino} A.,  {Hidalgo} S.~L.,  {Aparicio} A.,  {Gallart} C.,  {Carrera} R.,
  {Monelli} M.,  {Buonanno} R.,   {Marconi} G.,  2013, \mn@doi [\mnras]
  {10.1093/mnras/stt833}, \href
  {http://adsabs.harvard.edu/abs/2013MNRAS.433.1505D} {433, 1505}

\bibitem[\protect\citeauthoryear{{van den Bergh}}{{van den
  Bergh}}{1998}]{vdb98}
{van den Bergh} S.,  1998, \mn@doi [\apjl] {10.1086/311624}, \href
  {http://adsabs.harvard.edu/abs/1998ApJ...505L.127V} {505, L127}

\makeatother
\end{thebibliography}

\appendix

\section{Comparing the dynamical buoyancy effect from semi-analytical model to simulations}\label{app_sim}
\begin{figure*}
\begin{center}
\includegraphics[width=0.9\textwidth,trim=0 380 0 0,clip=true]{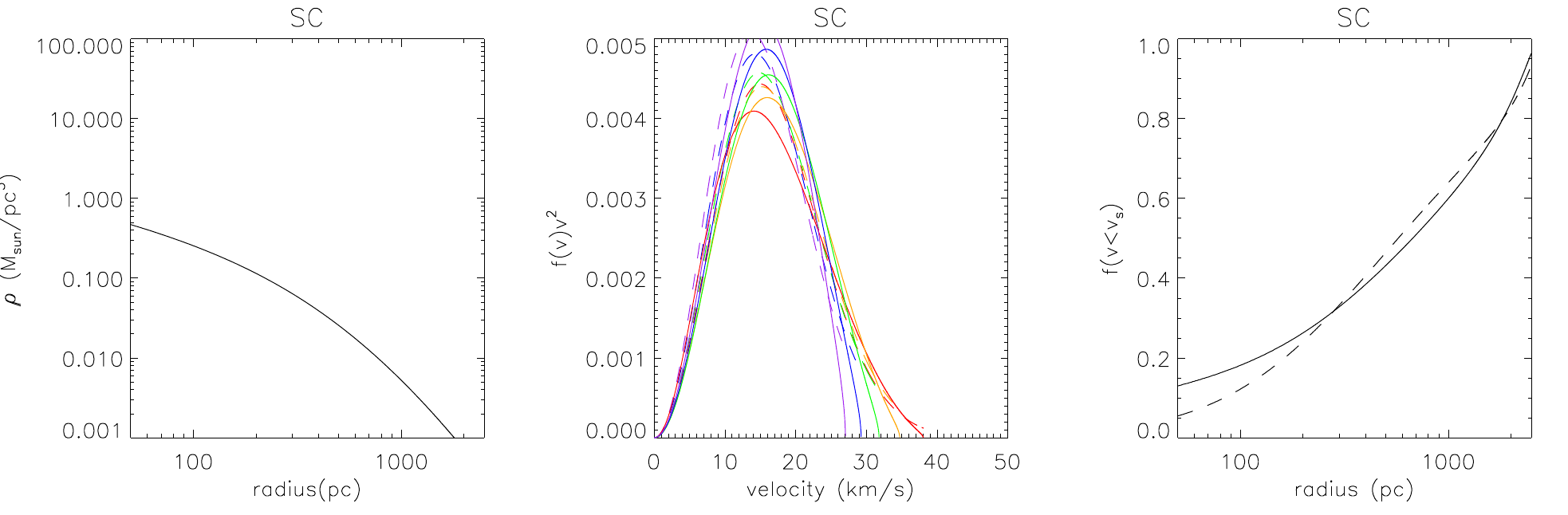}
\includegraphics[width=0.9\textwidth,trim=0 380 0 0,clip=true]{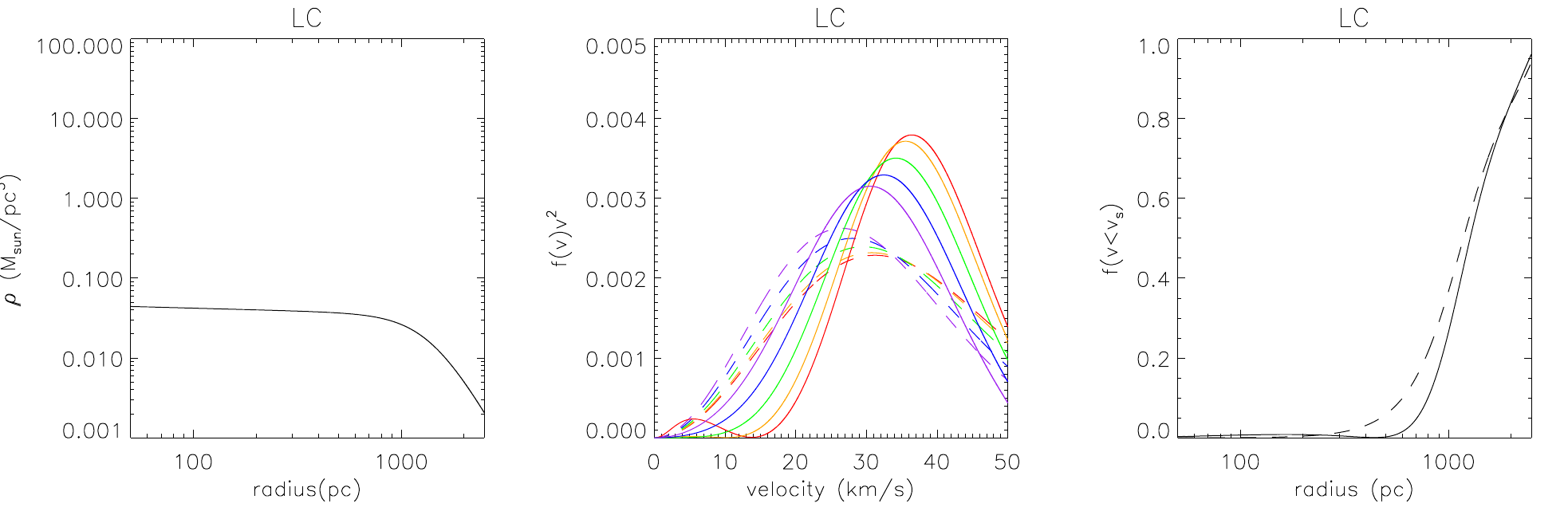}
\includegraphics[width=0.9\textwidth,trim=0 380 0 0,clip=true]{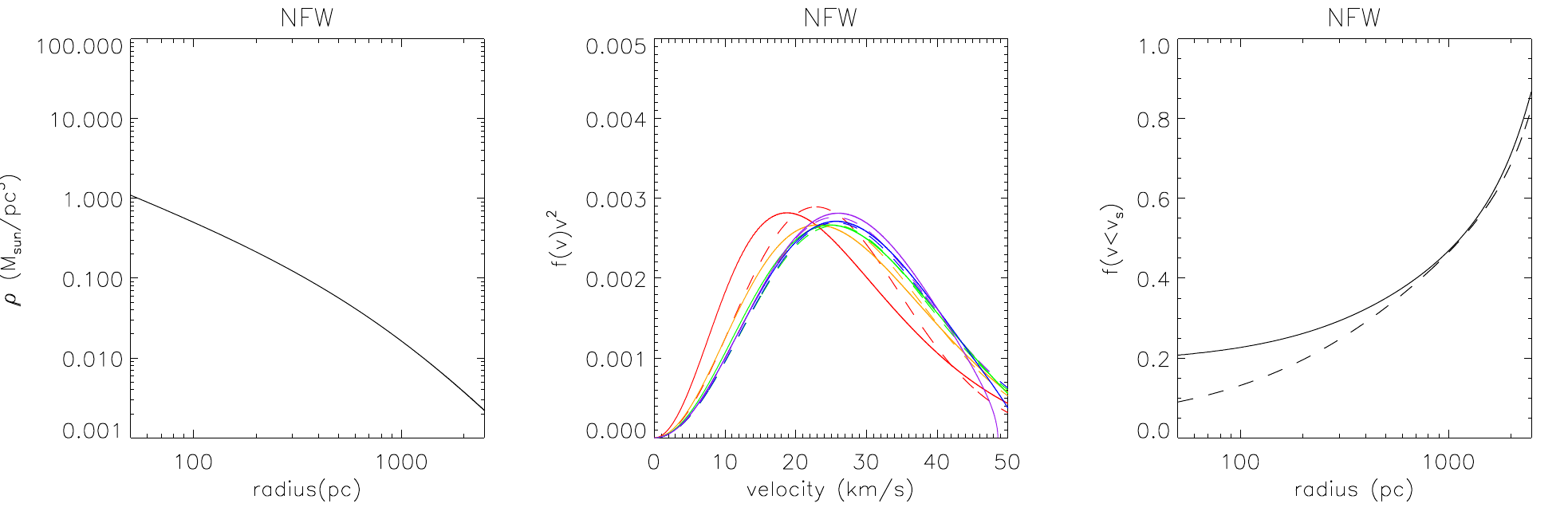}
\includegraphics[width=0.9\textwidth,trim=0 380 0 0,clip=true]{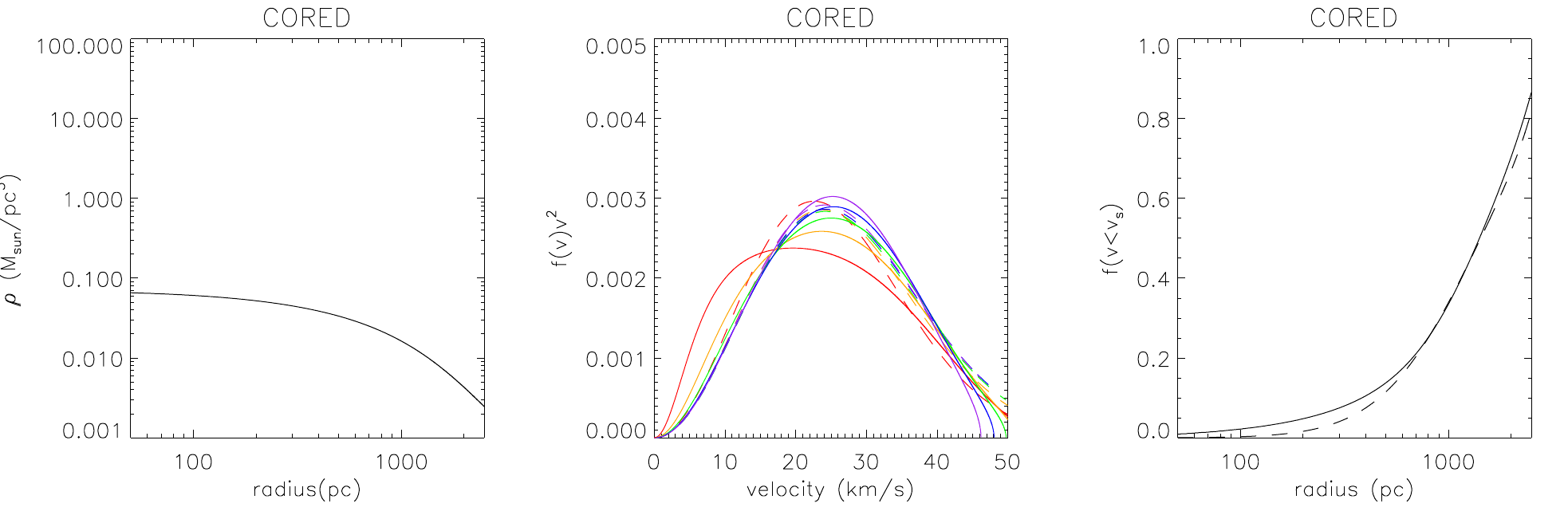}
\caption{Left and middle columns show the density profiles $\rho(r)$ and velocity distribution functions $f(v)$ derived from Eddington equation (solid) and assumed as a Maxwellian (dashed) distribution, from top to bottom, for the `SC', `LC', \texttt{nfw0} and \texttt{cored0} profiles. The different colours in the middle panels represent the velocity distribution functions evaluated at radii of, red: 200\,pc, orange: 400\,pc, green: 600\,pc, blue: 800\,pc, and purple: 1000\,pc.The velocity distribution function estimated with a Maxwellian assumption is increasingly erroneous towards the small galactic radii. In general, the Maxwellian assumption underestimates the fraction of slow background particles ($f(v<v_{s})$), as shown on the rightmost column.}
\label{fig_vdf}
\end{center}
\end{figure*}

Dynamical buoyancy was first demonstrated in $N$-body simulations done by \citet{cole12}. \citet{cole12} has run $N$-body simulations on the orbital decay of the five GCs in Fornax dSph under four different dark matter halo profiles which are labelled as strong-cusp (SC), intermediate-cusp (IC), weak-cusp (WC) and large-core (LC), and are progressively less cuspy in the order listed here. The details of these four profiles can be found in \citet{cole12}. Here we compare the stalling position of the GCs in the two extreme cases: SC and LC profiles, obtained from our analytical dynamical friction implementation with the ones obtained from $N$-body simulations by \citet{cole12}. Figure \ref{fig_vdf} shows the velocity distribution function $f(v_\bullet)$ of the dark matter particles of the SC and LC profiles tested in \citet{cole12}, as well as the \texttt{nfw0} and \texttt{cored0} profiles described in the main text. Cuspy profiles such as the SC and the \texttt{nfw0} profiles have larger fraction of slow particles (with $v_\bullet<v_\mathrm{s}$, here $v_\mathrm{s}$ is the velocity of the infalling satellite and is taken to be the circular velocity of the considered dark matter halo) than the cored profiles such as the LC and the \texttt{cored0} profiles.

\begin{figure}
\begin{center}
\includegraphics[width=0.5\textwidth,trim=0 0 0 0,clip=true]{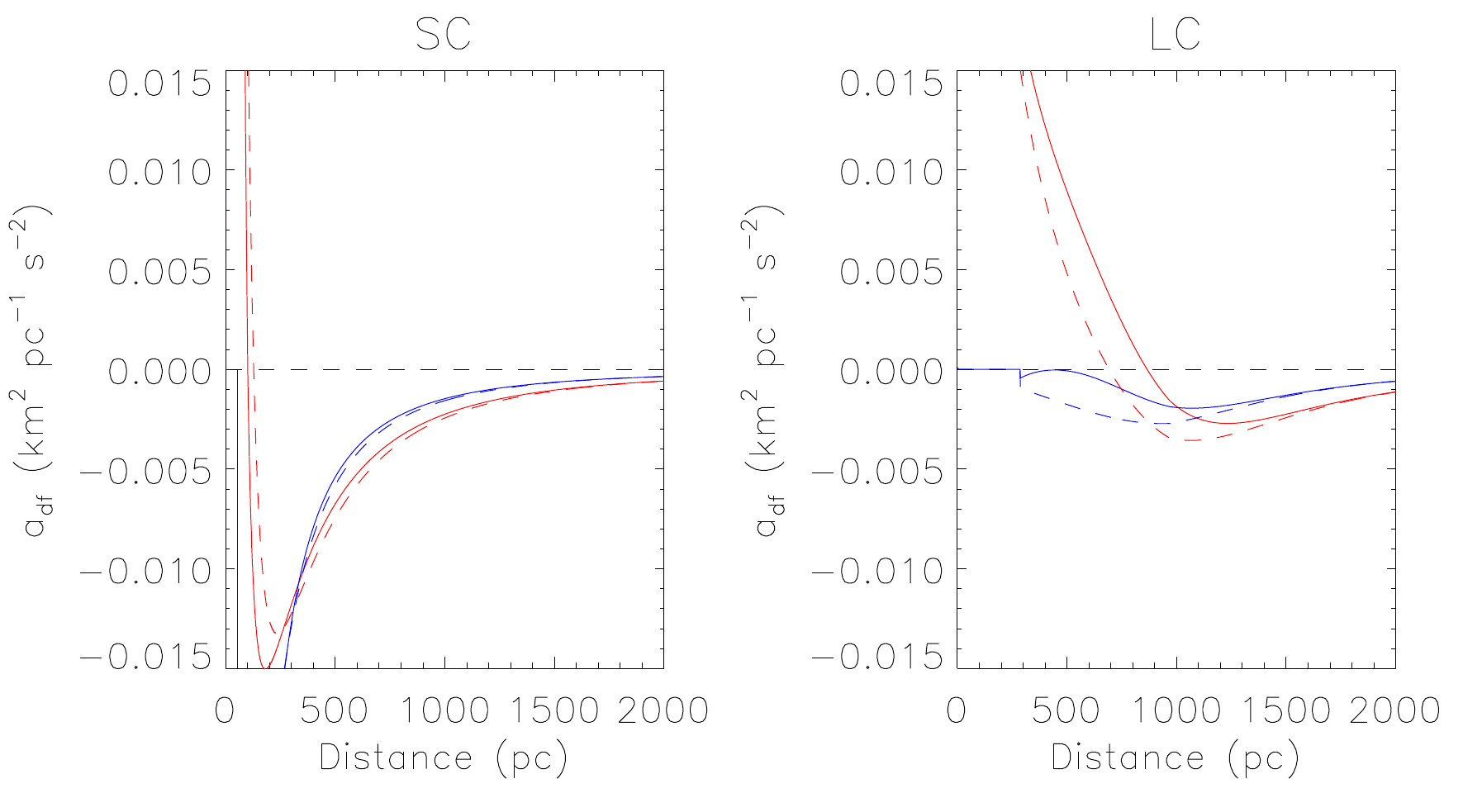}
\caption{Acceleration due to dynamical friction, $a_\mathrm{df}$, experienced by GC3 under the steep-cusp (SC) and large-core (LC) profiles in \citet{cole12}. Red lines denote dynamical friction treatments including fast stars (FS). Blue lines denote slow stars only (SS). Solid and dashed lines are runs using velocity distribution function from the Eddington equation (EDD) and Maxwellian assumptions (MAX) respectively. Horizontal dashed line represent where GC3 should stall as the dynamical friction and dynamical buoyancy balance out each other. Corresponding figure for the \texttt{nfw0} and \texttt{cored0} profiles can be found in Figure \ref{fig_adf}.}
\label{fig_adf_cole}
\end{center}
\end{figure}

Figure \ref{fig_adf_cole} shows the dynamical effects exerted on GC3 by the background dark matter particles for the SC and LC profiles. Just like Figure \ref{fig_adf}, GC3 experience dynamical friction at radii with $a_\mathrm{df}<0$ and dynamical buoyancy at radii with $a_\mathrm{df}>0$. The stalling radius is the radius at which $a_\mathrm{df}=0$. \citet{cole12} has shown that the stalling radius of LC is at $\sim$800\,pc, below which the GCs experience dynamical buoyancy.  This result is well-reproduced by our analytical model that include the effects of fast-moving background particles and a velocity distribution calculated from the Eddington equation, as shown in the red solid line on the right panel of Figure \ref{fig_adf_cole}. The stalling radius of LC is underestimated by $\sim25\%$ if we assume Maxwellian velocity distribution (red-dashed curve) and the dynamical buoyancy cannot be reproduced at all if we only consider slow-moving background particles (blue curves). We note that for the SC profile, our analytical model has predicted a stalling radius of $\sim100$\,pc, while the simulations of \citet{cole12} suggest that the GCs can sink below 10\,pc. Even without the inclusion of fast-moving particles, tidal stalling alone predicts a stalling radii of $\sim50$\,pc (blue curves). The discrepancies between analytic description of dynamical friction and the simulations could be caused by the lack of spatial resolution of the simulations, or that the velocity distributions in the innermost part of the halo cannot be captured by our simple assumptions. In either case, since the present-day location of the GCs are much greater than 100\,pc, especially for GC1, GC2, GC3 and GC5, which provide the strongest constraints on the halo profile and merger history, such discrepancies at small radii would not affect our results.

\section{Chemical support for the merger scenario from Fornax's chemical evolution}\label{subsect_mdf}
We consider here whether a merger with mass ratios of 1:5 to 1:2 are supported (or even permitted), given the observed metallically distribution function (MDF) of Fornax's field stars. For this exercise, we take the observed metallicity measurements of individual RGB stars within several local group dwarf galaxies \citep[][and references therein, as recalibrated by \cite{star10}]{lea13}, perform superpositions of pairs of dwarf galaxies and then compare the combined metallicity distribution with that of Fornax. To avoid possible systematics introduced by binning, we apply this analysis on the cumulative distribution function (CDF) instead of the MDFs themselves.

\begin{figure}
\begin{center}
\includegraphics[width=0.5\textwidth, trim= 45 350 20 80, clip = True ]{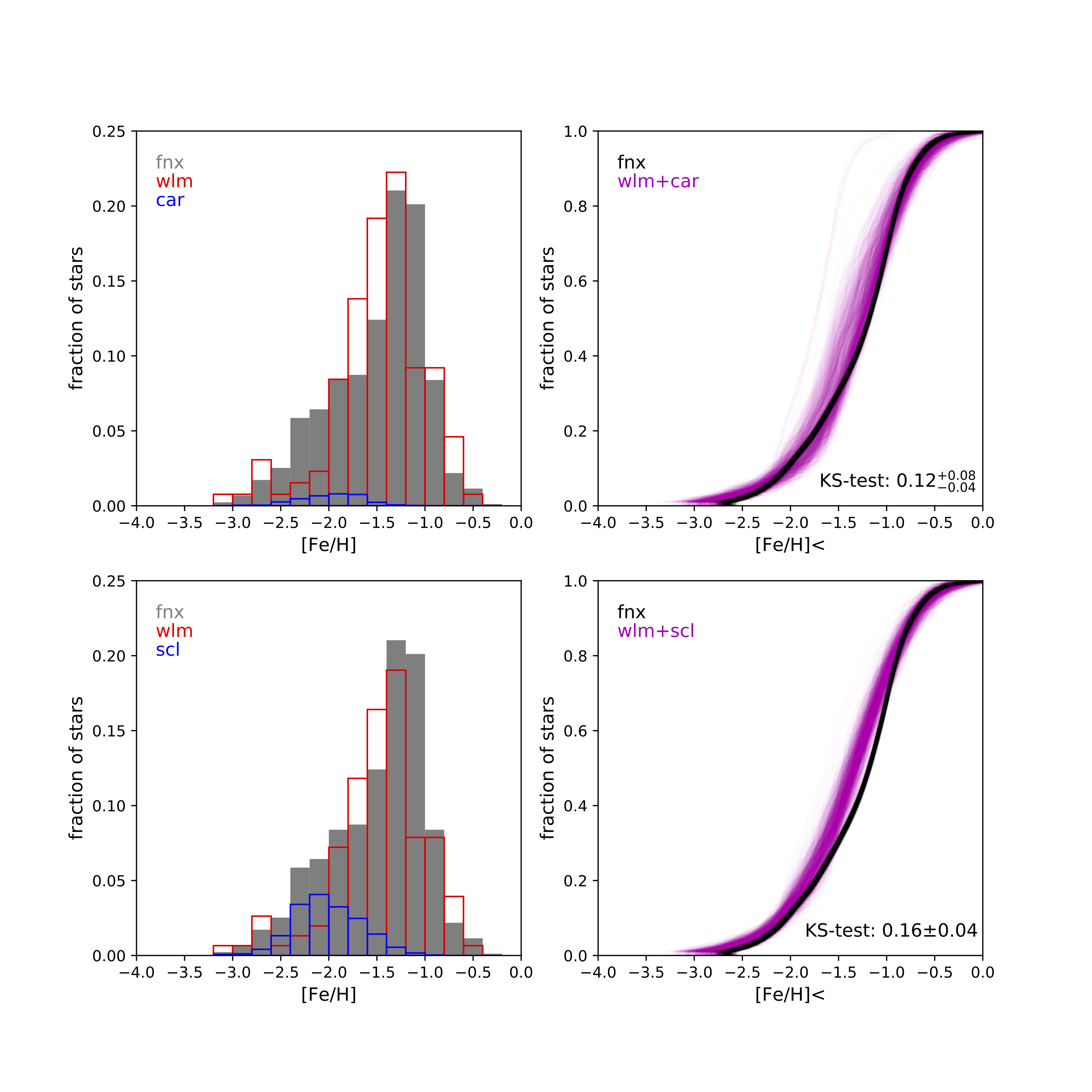}
\caption{Left: Normalised MDF of Fornax in grey, the mass-weighted MDFs of WLM and Carina in red and blue respectively. Right: 1000 Monte-Carlo realisations of the CDF of Fornax in black and that of the WLM and Carina combined in magenta. The mean and 1-$\sigma$ KS-test values are show in the bottom right corner.
\label{fig_carwlm}}
\end{center}
\end{figure}

\begin{figure}
\begin{center}
\includegraphics[width=0.5\textwidth, trim= 45 55 20 375, clip = True ]{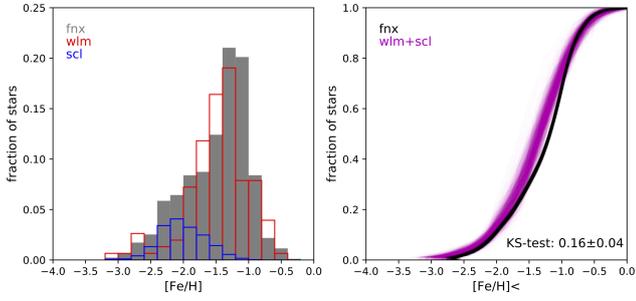}
\caption{Same as Figure \ref{fig_carwlm}, but for WLM and Sculptor.
\label{fig_sclwlm}}
\end{center}
\end{figure}

To demonstrate the feasibility and support the premise of a past merger for Fornax, we show in Figure \ref{fig_carwlm} and \ref{fig_sclwlm} the combined stellar metallicity CDF for two sets of galaxy pairs which satisfy the mass ratio requirements: WLM+Carina and WLM+Sculptor. With stellar masses of  1.1$\times$10$^{7}\,M_\odot$ \citep[WLM;][]{jack07}, 3.8$\times$10$^{5}\,M_\odot$ \citep[Carina;][]{mc12} and 2.3$\times$10$^{6}\,M_\odot$ \citep[Sculptor;][]{mc12}, a merger between WLM+Carina and WLM+Sculptor would constitute a 1:5 and a 1:2 merger respectively if we consider the stellar-mass-halo-mass (SMHM) relation from \citet{mos10} at redshift zero. 

We show the observed MDF of Fornax in grey in the left panel of Figure \ref{fig_carwlm}, and those of WLM and Carina in red and blue respectively. When computing the CDFs of each galaxy, we also take into account the measurement errors of the metallicities of individual stars. We did a Monte-Carlo sampling with 1000 realisations, each time varying the metallicity of each stars within a gaussian distribution with width equal to the star's measurement error.

We construct the combined CDF of Carina and WLM by drawing $N_1$ and $N_2$ stars from the normalised CDFs of the two galaxies, where $N_1$ and $N_2$ are determined by the stellar mass ratio between the two galaxies and $N_\mathrm{tot}=N_{1}+N_{2}$ is constrained by the total number of stars to be equal to the number of stars measured in Fornax. We again do a Monte-Carlo sampling with 1000 realisation, each time varying the total stellar mass within the measurement error, which we take to be 30\% of the measured value. 

The resultant 1000 realisations of the combined CDF is plotted in magenta and that of Fornax is plotted in black on the right panel of Figure \ref{fig_carwlm}. We next perform a Kolmogorov-Smirnov test between the combined WLM + Carina, and the Fornax metallicity CDFs for each realisation -- deriving a KS-test value of $0.12^{+0.08}_{-0.04}$. 

As a plausible representation of a dwarf-dwarf merger with mass ratio of 1:2, we compute a similar CDF of Sculptor and WLM as a comparison. We plot the MDF of Sculptor in blue on the left panel of Figure \ref{fig_sclwlm}. We perform the same exercise as in Figure \ref{fig_carwlm} to obtain a combined CDF of Sculptor and WLM, with the Monte-Carlo realisations of the combined CDF shown in magenta on the right panel of Figure \ref{fig_sclwlm}. The analysis of the simulated MDFs of these mergers show comparable K-S statistics within the uncertainties. Our exercise, while simple, gives independent support from empirical chemical properties that a merger with mass ratio anywhere between 1:5 to 1:2 could have happened in the past of Fornax, plausibly giving rise to its field star MDF.

\bsp

\label{lastpage}

\end{document}